\documentclass[preprint,10pt]{elsarticle}

\makeatletter
\def\ps@pprintTitle{%
 \let\@oddhead\@empty
 \let\@evenhead\@empty
 \def\@oddfoot{\centerline{\thepage}}%
 \let\@evenfoot\@oddfoot}
\makeatother

\journal{Physica A}
\usepackage{setspace}
\usepackage{amssymb} 
\usepackage{amsmath}
\usepackage{mathrsfs}
\usepackage{stmaryrd}
\usepackage{amsthm}
\usepackage{mathtools}
\usepackage{amsfonts}
\usepackage{comment}
\usepackage[colorinlistoftodos]{todonotes}
\usepackage{tikz-cd}   
\usepackage[utf8]{inputenc} 
\usepackage[T1]{fontenc}    
\usepackage{hyperref}       
\usepackage{url}            
\usepackage{nicefrac}       
\usepackage{microtype}      
\usepackage[utf8]{inputenc} 
\usepackage{booktabs}       
\usepackage{lingmacros}
\usepackage{tree-dvips} 
\usepackage{nccmath}
\usepackage{bbm}
\DeclareMathOperator*{\argmax}{argmax}
\usepackage{amsmath, amssymb, mathtools, bm, etoolbox}
\usepackage[noend]{algpseudocode}
\usepackage{algorithm}
\usepackage{mathrsfs}
\usepackage{caption}
\usepackage{graphicx}
\usepackage{subcaption}
\usepackage{color}
\usepackage{enumitem}

\newtheorem{thm-defn}[theorem]{Theorem/Definition}

\theoremstyle{definition}

\theoremstyle{remark}

\DeclareMathOperator{\E}{\mathbb{E}}

\newcommand{\ignore}[1]{}{}
\usepackage{lineno}
\begin{document}
\begin{frontmatter}
\title{Changes to the extreme and erratic behaviour of cryptocurrencies during COVID-19}
\author[label1]{Nick James} 
\author[label2]{Max Menzies} \ead{max.menzies@alumni.harvard.edu}
\author[label1]{Jennifer Chan} 
\address[label1]{School of Mathematics and Statistics, University of Sydney, NSW, Australia}
\address[label2]{Yau Mathematical Sciences Center, Tsinghua University, Beijing, China}

\begin{abstract}
This paper introduces new methods for analysing the extreme and erratic behaviour of time series to evaluate the impact of COVID-19 on cryptocurrency market dynamics. Across 51 cryptocurrencies, we examine extreme behaviour through a study of distribution extremities, and erratic behaviour through structural breaks. First, we analyse the structure of the market as a whole and observe a reduction in self-similarity as a result of COVID-19, particularly with respect to structural breaks in variance. Second, we compare and contrast these two behaviours, and identify individual anomalous cryptocurrencies. Tether (USDT) and TrueUSD (TUSD) are consistent outliers with respect to their returns, while Holo (HOT), NEXO (NEXO), Maker (MKR) and NEM (XEM) are frequently observed as anomalous with respect to both behaviours and time. Even among a market known as consistently volatile, this identifies individual cryptocurrencies that behave most irregularly in their extreme and erratic behaviour and shows these were more affected during the COVID-19 market crisis.




\end{abstract}

\end{frontmatter}
\linenumbers
\begin{nolinenumbers}
\section{Introduction}

The COVID-19 pandemic has had significant impacts on society, and prompted a substantial amount of attention and research. In epidemiology, studies have focused on the spread of COVID-19 and potential measures of containment \cite{Wang2020, Chinazzi2020, Liu2020, Fang2020, Zhou2020, Dehning2020}, while clinical research has explored potential treatments for the disease \cite{Jiang2020, Zu2020, Li2020, Zhang2020, Wang2020_CELL, NEJM2020}, including a vaccine \cite{Corey2020}. 
Researchers in nonlinear dynamics have used a wide range of new techniques to analyse and predict the spread of COVID-19 cases and deaths \cite{Khajanchi2020, Manchein2020, Ribeiro2020, Chakraborty2020, Anastassopoulou2020,James_Chaos}.

By contrast, most research studying the impact of COVID-19 on financial markets has used traditional statistical methods such as parametric models
\cite{Zhang2020finance,He2020, Zaremba2020, Mnif2020}. For example, \cite{Akhtaruzzaman2020, Akhtaruzzaman2020_2} use dynamic conditional correlations between stock returns and hedging costs to study financial contagion and safe asset classes. Similarly, \cite{Okorie2020} studies financial contagion via cross-correlation analysis, \cite{Corbet2020} uses a GARCH(1,1) model, while \cite{Yarovaya2020} is a qualitative overview of financial contagion. \cite{Conlon2020,Conlon2020_2} use a value at risk measure to study safe havens in the cryptocurrency market. \cite{Lahmiri2020} uses existing metrics and tests (such as approximate entropy, Lyapunov Exponents, t-tests and F-tests) to study the predictability of price fluctuations. All these methods are well-studied in the existing finance literature. Notably, \cite{Ji2020} proposes new methods to detect asset tail risk and determine if such risk can be reduced with safe asset classes.

Since their inception, cryptocurrencies have been of great interest to researchers in dynamical systems and econophysics. There has been substantial research on individual cryptocurrencies such as Bitcoin \cite{Chu2015, Lahmiri2018, Kondor2014, Bariviera2017, AlvarezRamirez2018} and the disorder and fractal behaviour within cryptocurrencies in general \cite{Stosic2019,Stosic2019_2, Manavi2020, Ferreira2020}. Notably, Drozdz and coauthors have analysed several quantitative aspects of the cryptocurrency market closely, including cross-correlations, auto-correlations and scaling effects. Across three papers \cite{Drod2018, Drod2019, Drod2020} they have proposed significant evidence that the cryptocurrency market has become an independent regular market decoupled from and resembling foreign exchange. Structural breaks of cryptocurrencies have been analysed by \cite{Telli2020,James2020_nsm}, who imply, but do not explicitly state, that these points in time where statistical properties change herald erratic and unpredictable behaviour.

The goal of this paper is to extend the study of the nonlinear dynamics of cryptocurrencies and introduce new methods to analyse their extreme and erratic behaviour, before and during the COVID-19 pandemic. Extreme behaviour is analysed via restricted distributions that capture the extreme values of the log returns and Parkinson variance time series; erratic behaviour is analysed via structural breaks. We build on the existing literature in several ways. The theory of extreme events has been studied in \cite{Qi2019} and \cite{Yang2018,Yang2019} where machine learning methods have been used to predict extreme events and detect outliers, respectively. While \cite{Conlon2020} and \cite{Mnif2020} study the impact of COVID-19 on 3 and 5 cryptocurrencies, respectively, and \cite{Telli2020} studies the structural breaks of 7 cryptocurrencies, we analyse the impact of COVID-19 on the extreme behaviour and structural breaks of 51 cryptocurrencies. While \cite{Akhtaruzzaman2020,Akhtaruzzaman2020_2, Okorie2020, Corbet2020,Conlon2020,Conlon2020_2,Lahmiri2020} have studied the financial impact of the COVID-19 pandemic with existing techniques, all our techniques are new. Finally, we develop a more general framework than that proposed by \cite{Ji2020}. We determine changes due to market dynamics in general, and identify several specific cryptocurrencies that are anomalous with respect to returns or variance. Further, we develop new \emph{inconsistency matrices} to compare the relationship between extreme and erratic behaviours before and after the emergence of COVID-19.

In Section \ref{sec:methodology}, we describe our methodology to analyse extreme and erratic behaviour of an arbitrary collection of time series. The methods therein may be applied more generally than in the instance of this paper. In Section \ref{sec:results}, we apply our methods to the log returns and Parkinson variance \cite{Parkinson1980}  time series for 51 cryptocurrencies. We conclude in Section \ref{sec:conclusion}.

\section{Methodology}
\label{sec:methodology}
In this paper, the most general object of study is a collection of real-valued time series $X_t^{(i)}, i=1,...,n$ over a time interval $t=1,...,T$. We analyse four such collections: the log returns and Parkinson variance of 51 cryptocurrencies before and during the COVID-19 pandemic. Let $P_t,H_t,L_t$ be the closing price, the daily high, and the daily low, respectively, of a financial instrument at time $t$. Let $R_t$ and $\sigma^2_t$ be the log returns and Parkinson variance time series, respectively, defined as follows:
\begin{align}
\label{eq:logreturns}    R_{t} &= \log \left(\frac{P_{t}}{P_{t-1}}\right),\\
  \label{eq:parkvar}  \sigma^2_t&=\frac{(\log H_{t}-\log L_{t})^2}{4 \log 2}.
\end{align}
The Parkinson variance $\sigma^2_t$ defined above is a measure of the instantaneous price variance of a financial asset based on its intra-day volatility \cite{Parkinson1980}. This gives a time-dependent variance rather than a scalar value through the standard deviation computation. The time series $R_t$ takes both positive and negative values, while $\sigma^2_t$ is non-negative everywhere, a distinction that is necessary in Section \ref{methodology:extreme}. The precise methodology that we describe below is not exhaustive. Below we outline the detailed implementation to model the erratic and extreme behaviour of cryptocurrencies. Modifications may be made for other contexts.

\subsection{Distance between distribution extremities}
\label{methodology:extreme}
In this section, we describe how we extract and measure the distance between the extreme values of various time series. Let $\mu$ be a probability distribution that records the values of a time series $X_t$. For full generality, suppose $\mu$ is a continuous probability measure of the form $\mu=f(x)dx$, where $dx$ is Lebesgue measure, and $f(x)$ a probability density function. As such, $f(x)$ is non-negative everywhere with integral $1$.

For the log return time series $R_t$, we extract the points of density $5$\% and $95$\%, respectively, by the equations

\begin{align}
\int_{-\infty}^s f(x) dx &= 0.05 \\
\int_{t}^\infty f(x) dx &= 0.05
\end{align}
The range $x\leq s$ gives the left extremal 5\% of the distribution, while the range $x\geq t$ gives the right extremal 5\%. Next, we define the restricted function by 
\begin{align}
g(x)=f(x) \mathbbm{1}_{\{x \leq s\}\cup \{x\geq t\}}=
\begin{cases}
f(x), x \leq s \\
0, s<x<t \\
f(x), x \geq t.
\end{cases}
\end{align}
Above, $\mathbbm{1}$ denotes an indicator function of a set; this construction essentially truncates $f$ only in its tail range. Next, we form the associated measure $\nu=g(x) dx$, where $dx$ is Lebesgue measure. This construction is common in probability theory \cite{Billingsley}.

\vspace{1em}

For the Parkinson variance time series $\sigma^2_t$, the associated function $f$ is supported on $(0,\infty).$ In this case, we extract the point of density 90\% by 
\begin{align}
\int^{\infty}_{l} f(x) dx = 0.10.
\end{align}
In this instance, we truncate $f$ only in its extremal positive range by defining 
\begin{align*}
h(x)=f(x) \mathbbm{1}_{\{x\geq l\}}= 
\begin{cases}
f(x), x \geq l \\
0, x < l.
\end{cases}
\end{align*}
Again, we form the associated measure $\eta = h(x) dx$ where $dx$ is Lebesgue measure. In both cases, this procedure works even more simply for a discrete distribution given by a finite data set. For the log returns, we form the empirical distribution function, then remove the middle $90$\% of the values by order; for the Parkinson variance, we remove the bottom 90\% of the values.

Now suppose we are given $n$ time series $X_t^{(i)}, i=1,...,n$. We form $n$ associated probability measures $\mu_1,...,\mu_n$ and then the restricted measures $\nu_1,...,\nu_n$ for the log returns and $\eta_1,...,\eta_n$ for the Parkinson variance. All these restricted measures have total measure equal to 0.1 so we may compute the $L_1$-Wasserstein metric $d_1$ \cite{delBarrio} between them. Also known as the Earth-mover's distance, this is a common metric between two measures with the same total mass. The construction of the ``associated measure'' of $\nu$ and $\eta$ is a technical step from probability theory \cite{Billingsley} that means the Wasserstein metric can compute distance between two distribution functions by transforming them into measures. We conclude by forming a matrix between the distribution extremities of the time series. Let $D^{ER}_{ij}=d_1(\nu_i,\nu_j)$ be the matrix between the log return distributions, and $D^{EV}_{ij}=d_1(\eta_i,\eta_j)$ be the matrix between the Parkinson variance distributions.

In Section \ref{sec:results}, we will consider the means $\E(\nu_i)$ of the restricted distributions associated to the log returns. Individually, $\E(\nu_i)>0$ if the corresponding cryptocurrency has greater extreme positive than negative returns, on average. Collectively, they will offer us insight into the market as a whole.

Altogether, studying the log returns, Parkinson variance time series and their tail ends gives us insight into not only the mean and standard deviation of the cryptocurrencies but also higher moments from studying the tail behaviour. In particular, applying our method to the time-dependent Parkinson variance allows us to study extremities in variance in detail.

\subsection{Distance between erratic behaviour profiles}
\label{metholody:erratic}
Let $X_t$ be a time series, $t=1,...,T$. We apply the two-phase \emph{change point detection algorithm} described by \cite{Ross2015} to obtain a set of structural breaks, also known as change points. These are points in time where the statistical properties of the time series change, as detected by a particular algorithm. Applied to a collection $X_t^{(i)},i=1,...,n, t=1,...,T$ of time series, this produces a collection of finite sets $S_1,...,S_n$, each a subset of $\{1,...,T\}$. Further details are given in \ref{Appendix_CPD}. This describes a specific change point detection algorithm. Our methodology is flexible and may build off any such algorithm.

Next, we measure appropriate distances between the sets $S_i$. Traditional metrics such as the Hausdorff distance are unsuitable, being too sensitive to outliers \cite{Baddeley1992} so we adopt and modify the semi-metrics developed in \cite{James2020_nsm} between the sets of structural breaks $S_i$. We define a normalised distance as follows:
\begin{equation}
    D({S_i},{S_j}) = \frac{1}{2T} \Bigg(\frac{\sum_{b\in S_j} d(b,S_i)}{|S_j|} + \frac{\sum_{a \in {S_i}} d(a,S_j)}{|S_i|} \Bigg),
\end{equation}
where $d(b,S_i)$ is the minimal distance from $b \in S_j$ to the set $S_i$. The expression above is a $L^1$ norm average of all minimal distances from elements of $S_1$ to $S_2$ and vice versa, normalised by both the size of the sets and the length of the time series. As in Section \ref{methodology:extreme}, we form a matrix between the sets of structural breaks, $D^B_{ij}=D(S_i,S_j)$. Let $D^{BR}$ and $D^{BV}$ be the matrices between sets of structural breaks for the log return and Parkinson variance time series, respectively.

\subsection{Time-varying dynamics of the cryptocurrency market}
\label{methodology:timevarying}
In this section, we describe how we analyse the dynamics of returns and variance across the entire cryptocurrency market, before and after COVID-19. Let the Frobenius norm of a vector $v \in \mathbb{R}^n$ and an $n \times n$ matrix $D$ be defined as $||v||=\left(\sum_{i=1}^n |v_i|^2 \right)$ and $||D||=\left(\sum_{i,j=1}^n |d_{ij}|^2\right) ^{\frac{1}{2}}$, respectively. Let $\mathbf{R}_t$ and $\Sigma_t$ be the length $n$ vectors of all log returns $(R_t^{(i)})$ and Parkinson variances $((\sigma^2_t)^{(i)})$ at time $t$, respectively. The values $||\mathbf{R}_t||$ and $||\Sigma_t||$ give the magnitude of the first two distribution moments across the market, and represent its volatility as a whole. The Frobenius norm of an $n \times n$ distance matrix $D$ represents the total size of all distances within a collection of $n$ elements. A greater Frobenius norm indicates less self-similarity in the collection. Having studied overall market dynamics, we next seek to understand the changing relationships between cryptocurrencies. For this purpose, we study four different time series consisting of the log returns and Parkinson variance over two different periods, and compare the distance matrices we have defined, pertaining to return extremes, variance extremes, return breaks and variance breaks. We compute the respective matrix norms before and after the emergence of COVID-19 to understand the pandemic's impact on the similarity and dynamics of the cryptocurrency market. Our periods of analysis are 30-06-2018 to 31-12-2019 as ``pre-COVID,'' which we denote with the subscript ``pre'' and 1-1-2020 to 24-06-2020 as ``post-COVID,'' which we denote with the subscript ``post.'' Our methods may appropriately compare periods of different length as all matrices are appropriately normalised. Thus, we have eight different distance matrices:
\begin{itemize}
    \item For the log returns pre-COVID time series, we have $D_{\text{pre}}^{ER}$ and $D_{\text{pre}}^{BR}$;
    \item For the Parkinson variance pre-COVID time series, have $D_{\text{pre}}^{EV}$ and $D_{\text{pre}}^{BV}$;
    \item For the log returns post-COVID time series, have $D_{\text{post}}^{ER}$ and $D_{\text{post}}^{BR}$;
    \item For the Parkinson variance post-COVID time series, have $D_{\text{post}}^{EV}$ and $D_{\text{post}}^{BV}$.
\end{itemize}

\noindent We compute the Frobenius norms for these eight matrices in Section \ref{sec:results}.

\subsection{Inconsistency analysis}
\label{methodology:crosscontext}
In this section, we describe how we measure the consistency between the extreme and erratic behaviours of cryptocurrencies. To do so, we introduce a new method of comparing distance matrices and apply this to compare $D^B$ and $D^E$ for both the log return and Parkinson variance time series.

Given an $n \times n$ distance matrix $A$, the \emph{affinity matrix} is defined as 
\begin{equation}
    A_{ij} = 1 - \frac{D_{ij}}{\max{\{D\}}}.
\end{equation}
All elements of these affinity matrices lie in $[0,1]$, so it is appropriate to compare them directly by taking their difference. Let the affinity matrices associated to $D^{ER}, D^{EV}$, defined in Section \ref{methodology:extreme}, and $D^{BR}, D^{BV}$, defined in Section \ref{metholody:erratic},  be $A^{ER},A^{EV},A^{BR}$ and $A^{BV}$ respectively. We define the \emph{behaviour inconsistency matrix} between extreme and erratic behaviour be defined as follows, for log returns and Parkinson variance, respectively:
\begin{align}
        \text{INC}^{R,EB} &= A^{ER} - A^{BR}\\
        \text{INC}^{V,EB} &= A^{EV} - A^{BV}.
\end{align}
As described in Section \ref{methodology:timevarying}, we analyse the log returns and Parkinson variances over two distinct periods (pre- and post-COVID), so we have four different behaviour inconsistency matrices $\text{INC}^{\text{pre},R}_{\text{bhvr}},\text{INC}^{\text{pre},V}_{\text{bhvr}}, \text{INC}^{\text{post},R}_{\text{bhvr}}$ and  $\text{INC}^{\text{post},V}_{\text{bhvr}}.$

Finally, we can also define \emph{time inconsistency matrices}. By comparing corresponding distances matrices, extreme or erratic, over time, we can continue the goal of Section \ref{methodology:timevarying} and identify individual cryptocurrencies that have significantly changed with respect to their similarity with others. We define four inconsistency matrices with respect to time as follows:
\begin{align}
        \text{INC}^{ER}_{\text{time}} &= A^{ER}_\text{pre} - A^{ER}_\text{post}\\
        \text{INC}^{EV}_{\text{time}} &= A^{EV}_\text{pre} - A^{EV}_\text{post}\\
            \text{INC}^{BR}_{\text{time}} &= A^{BR}_\text{pre} - A^{BR}_\text{post}\\
        \text{INC}^{BV}_{\text{time}} &= A^{BV}_\text{pre} - A^{BV}_\text{post}.
\end{align}
Thus, we have defined eight different inconsistency matrices above. Four, $\text{INC}^{\text{pre},R}_{\text{bhvr}},\text{INC}^{\text{pre},V}_{\text{bhvr}}, \text{INC}^{\text{post},R}_{\text{bhvr}}$ and  $\text{INC}^{\text{post},V}_{\text{bhvr}}$, we refer to as \emph{behaviour inconsistency matrices}; the other four, $\text{INC}^{ER}_{\text{time}}, \text{INC}^{EV}_{\text{time}}, \text{INC}^{BR}_{\text{time}}$ and $\text{INC}^{BR}_{\text{time}}$, we refer to as \emph{time inconsistency matrices}. The \emph{behaviour inconsistency matrices} reveal cryptocurrencies that are irregular with respect to the rest of the market in the comparison between extreme and erratic behaviour. This may pertain to the returns or variance, pre- or post-COVID, depending on which of the four behaviour inconsistency matrices we examine. As a shorthand, we will say such cryptocurrencies are \emph{inconsistent with respect to behaviour.} The \emph{time inconsistency matrices} reveal cryptocurrencies that have been anomalously impacted by the emergence of COVID-19. This may pertain to its extreme or erratic behaviour of its log returns or variance, depending on which of the four time inconsistency matrices we examine. We will say such cryptocurrencies are \emph{inconsistent with respect to time}.

We reveal inconsistent cryptocurrencies as follows. Assume the cryptocurrencies are given some ordering $j=1,...,n$. Given any inconsistency matrix $\text{INC}$ of the eight above, define the \emph{anomaly score} of the $j$-th cryptocurrency as $a_j = \sum_{j=1}^n |\text{INC}_{ij}|$. Larger values indicate cryptocurrencies that are more anomalous between the two affinity matrices under consideration (recall that every inconsistency matrix is the difference of two affinity matrices). In Section \ref{sec:results}, we highlight the top 3 cryptocurrencies across a range of these inconsistency matrices to determine the most prominent anomalies. We have chosen to list the top 3 for ease of interpreting the results, and to point out that examining the most inconsistent cryptocurrencies reveals a considerable number of repetitions - that is, certain cryptocurrencies that are inconsistent in one regard frequently are inconsistent in other regards.

\section{Experimental results and discussion}
\label{sec:results}
We draw data from Coinmarketcap. Of the cryptocurrencies with price histories that go as far back as 30-06-2018, we analyse the 51 largest by market capitalisation. These are all detailed in Table \ref{tab:CryptocurrencyTickers}, ordered by decreasing market capitalisation. For each cryptocurrency, we draw closing price, daily high, and daily low, and first calculate the log returns and Parkinson variance as defined in (\ref{eq:logreturns}) and (\ref{eq:parkvar}). In the proceeding sections, we report a broad range of findings on the contrasting impact COVID-19 has had on the extreme and erratic behaviour of cryptocurrency returns and variance, and the identification of inconsistencies in various cryptocurrency behaviours. We refer to the period from 30-06-2018 to 31-12-2019 as ``pre-COVID'' and the period from 1-1-2020 to 24-06-2020 as ``post-COVID.''

We have chosen these periods deliberately to study the impact of the COVID-19 epidemiological crisis on financial assets, specifically on the cryptocurrency market. The post-COVID period can be characterised as a systemic crisis across all asset classes, while the pre-COVID period was relatively stable, with consistently positive returns across equity markets in 2019. Market crises refer to market crashes that impact all asset classes rather than individual drops in assets, even if they are precipitous.

This is not to say that there was not some drawdown within the cryptocurrency market during 2019, nor that all volatility during the post-COVID period can be explained by the pandemic. 
Indeed, the crash of the BitMEX exchange market crash in March 2020 caused profound losses in Bitcoin \cite{Forbes}. However, the pre-COVID and post-COVID periods are distinct in nature - the latter is characterised by the emergence of COVID-19 around the world and with subsequent crises in various equities and economic measures. Generally speaking, we are interested in studying the impact of broader financial crises on cryptocurrencies.

The periods are of different length, however this is appropriate given that all constructions in Section \ref{sec:methodology} are normalised with respect to time. Moreover, such a difference in lengths between crisis and non-crisis periods is inevitable. Historically, financial markets have exhibited a tendency for collective asset prices to rise slowly during ``bull'' markets and drop more quickly during ``bear'' markets. For example, The Dow Jones Industrial Average exhibited losses of -22.61\% on 19-10-1987 (Black Monday), -7.87\% on 15-10-2008 (the Global Financial Crisis), and -9.99\% on 12-03-2020 (Black Thursday) \cite{CNBC}. Accordingly, we have chosen to study a shorter period of time as our crisis period to more accurately reflect the market dynamics during the crash.


\subsection{COVID-19 impact on market dynamics} 
In this first section, we study the cryptocurrency market dynamics over time as a whole, without reference to the internal similarity between distinct cryptocurrencies. We use the same 51 cryptocurrencies as elsewhere in the paper, listed in Table A.3. In Figures \ref{fig:Return_norm} and \ref{fig:Variance_norm}, we depict the Frobenius norms $||\mathbf{R}_{t}||$ and $||\Sigma_{t}||$, respectively, defined in Section \ref{methodology:timevarying}. Larger magnitudes indicate more unstable market dynamics, characterised by greater total returns and variance, regardless of direction. These figures reveal several insights: first, during 2018 and 2019, the Parkinson variance exhibits less regular peaks than the log returns, but relative to the baseline these peaks are more significant. These peaks also appear to occur more periodically than that of returns, with our measure possibly highlighting some latent volatility clustering. Second, the sharpest peak of each figure is coincident: anomalously large changes in returns are accompanied by proportionally large changes in variance at the same time. The day of greatest changes in returns and volatility is 12 March, 2020. Known as Black Thursday \cite{CNBC}, this was the worst crash of the Dow Jones since 1987, and caused substantial drops across all sectors. On the exact same day, the cryptocurrency exchange BitMEX crashed, causing a substantial drop in the price of Bitcoin. After this greatest spike, both measures drop to levels below their 30-month average. That is, after this point, the volatility of the market is in fact reduced relative to the previous year.

\begin{figure*}
    \centering
    \begin{subfigure}[b]{0.95\textwidth}
        \includegraphics[width=\textwidth]{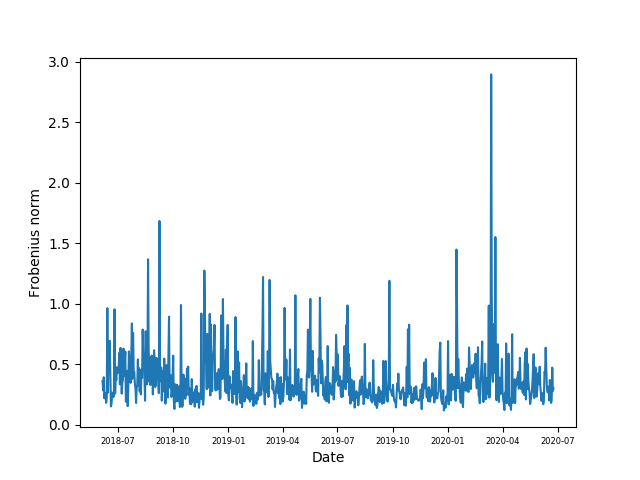}
        \caption{The Frobenius norm $||\mathbf{R}_{t}||$ as a function of time}
    \label{fig:Return_norm}
    \end{subfigure}
    \begin{subfigure}[b]{0.95\textwidth}
        \includegraphics[width=\textwidth]{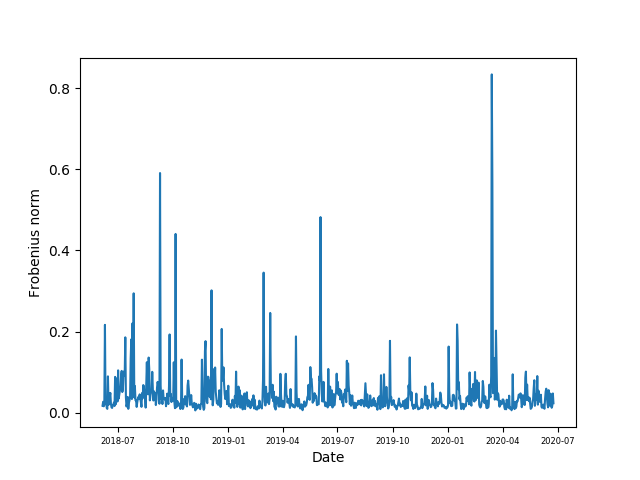}
        \caption{The Frobenius norm $||\Sigma_{t}||$ as a function of time}
    \label{fig:Variance_norm}
    \end{subfigure}
    \caption{The changing dynamics of the Frobenius norm for (a) log returns and (b) Parkinson variance over the whole market are plotted with time. A coincident sharp peak in March 2020 can be seen, showing the extreme but brief impact of COVID-19.}
    \label{fig:MarketDynamics}
\end{figure*}

\subsection{Changing dynamics of extreme and erratic behaviour}
In this section, we analyse the distance matrices between distribution extremities and structural breaks for the four time series under consideration, that is, log returns and Parkinson variance before and after the emergence of COVID-19. Specifically, we examine the Frobenius norms, defined in Section \ref{methodology:timevarying}, of the eight distance matrices defined in the same section. Greater Frobenius norms indicate greater overall distances and hence less self-similarity among a collection of cryptocurrencies. This reveals several insights:

First, both before and during COVID-19, the behaviour of cryptocurrencies is more self-similar (measured by Frobenius matrix norms) with respect to variance than returns. This is true for both distribution extremities and structural breaks, heralding extreme and erratic behaviour, respectively. This can be seen in the consistently smaller values in Tables  \ref{tab:extreme_value_norm_table} and \ref{tab:structural_break_norm_table}, where Frobenius norms, defined in Section \ref{methodology:timevarying}, are computed. All distances in question, and hence the matrix norms, are normalised, so this comparison is appropriate when comparing either distribution extremities or structural breaks among themselves. 

Next, the post-COVID period generally exhibits less similarity (measured by the Frobenius norms) in the cryptocurrency market than before COVID-19. Comparing distribution extremities and structural breaks with respect to the log returns and variance time series, three of these distance matrices exhibit greater Frobenius norm for the post-COVID period. Only the structural breaks in returns observe a slight decrease in Frobenius norm, implying greater self-similarity with respect to structural breaks of returns, in the post-COVID period. By contrast, the Frobenius norm for structural breaks in variance has increased almost threefold. Indeed, this is reflected in Figure \ref{fig:VolatilitybreaksDendrogram}. As we will discuss further in Section \ref{sec: breaks variance}, Figure \ref{fig:Volatility_breaks_dendrogram_pre} shows essentially one cluster, broad self-similarity and no outlier elements between structural breaks with respect to variance, while Figure \ref{fig:Volatility_breaks_dendrogram_post} has less self-similarity - rather, several anomalous elements are visible in the dendrogram.

On the other hand, for the Frobenius norms of distribution extremities, a moderate and similar increase was observed for both returns and variance. That is, the total similarity (measured by Frobenius norms) in return and variance extremes decreased by a similar amount due to COVID-19. The slight increase in Frobenius norm for distribution extremities of variance contrasts with the large increase in the case of structural breaks; while the increase in norms for log returns contrasts with the decrease in the case of breaks.

\begin{table}
\begin{tabular}{ |p{3.75cm}||p{2.95cm}|p{2.95cm}|}
 \hline
 \multicolumn{3}{|c|}{Matrix norms $||D^E||$ for returns and variance, pre- and post-COVID} \\
 \hline
 Period & Log returns & Parkinson variance \\
 \hline
Pre-COVID & 1.30 & 0.64   \\
Post-COVID & 1.51 & 0.73  \\
\hline
\end{tabular}
\caption{Frobenius norms for distribution extremity distance matrices pre- and post-COVID.}
\label{tab:extreme_value_norm_table}
\end{table}

\begin{table}
\begin{tabular}{ |p{3.75cm}||p{2.95cm}|p{2.95cm}|}
 \hline
 \multicolumn{3}{|c|}{Matrix norms $||D^B||$ for returns and variance, pre- and post-COVID} \\
 \hline
 Period & Log returns & Parkinson variance \\
 \hline
Pre-COVID & 5.70 & 0.84   \\
Post-COVID & 5.22 & 2.44  \\
\hline
\end{tabular}
\caption{Frobenius norms for structural breaks distance matrices pre- and post-COVID.}
\label{tab:structural_break_norm_table}
\end{table}

\subsection{Structural breaks with respect to Parkinson variance}
\label{sec: breaks variance}
In this section, we take a closer look at the structure of the cryptocurrency market with respect to structural breaks in Parkinson variance. In Figure \ref{fig:VolatilitybreaksDendrogram}, we depict the results of \emph{hierarchical clustering} of the corresponding affinity matrices $A_{\text{pre}}^{BV}$ and $A_{\text{post}}^{BV}$.   
This supports our analysis in the preceding section. There are two types of hierarchical clustering: agglomerative (bottom-up) and divisive (top-down) clustering. The former starts with each object defined as a single element cluster and sequentially combines the most similar clusters into a larger cluster, until there is only one cluster. The latter works by sequentially splitting the most heterogeneous cluster in two, until each cluster contains only a single element. Our experiments apply agglomerative hierarchical clustering. Unlike K-means clustering, hierarchical clustering does not require a specific number of clusters $k$ to be set, and the procedure generates an image of the cluster structure. Tree colors are determined by the dendrogram, which computes the optimal trade-off between the number of clusters and total error. An ideal fit would have fewer clusters and smaller error. More details are provided in \cite{mullner2011modern, Mllner2013}.

Figure \ref{fig:Volatility_breaks_dendrogram_pre} exhibits essentially one amorphous cluster lacking any visible structure, with no notable outliers and a high degree of self-similarity. The complete lack of subclusters represents the significant risk for investors in the cryptocurrency market even during normal times. Indeed, there is really no way to diversify against erratic behaviour in volatility, as all cryptocurrencies have highly similar structural breaks, and this exposes investors to simultaneous drawdown risk among any held collection of cryptocurrencies. After the emergence of COVID-19, we see a considerable change in market structure, displayed in Figure \ref{fig:Volatility_breaks_dendrogram_post}. There, two primary clusters of cryptocurrencies are observed. Within the smaller cluster, a subcluster of high similarity emerges containing cryptocurrencies such as Chainlink (LINK) and Tezos (XTZ). Several slight outliers emerge: Maker (MKR), Tether (USDT) and Zcash (ZEC). All cryptocurrency tickers are indexed in Table \ref{tab:CryptocurrencyTickers}. That is, the emergence of COVID-19 has significantly altered the market structure with respect to structural breaks in variance, from one highly homogeneous cluster with no outliers, to two clusters with some subcluster structure and slight outliers.

\begin{figure*}
    \centering
    \begin{subfigure}[b]{0.95\textwidth}
        \includegraphics[width=\textwidth]{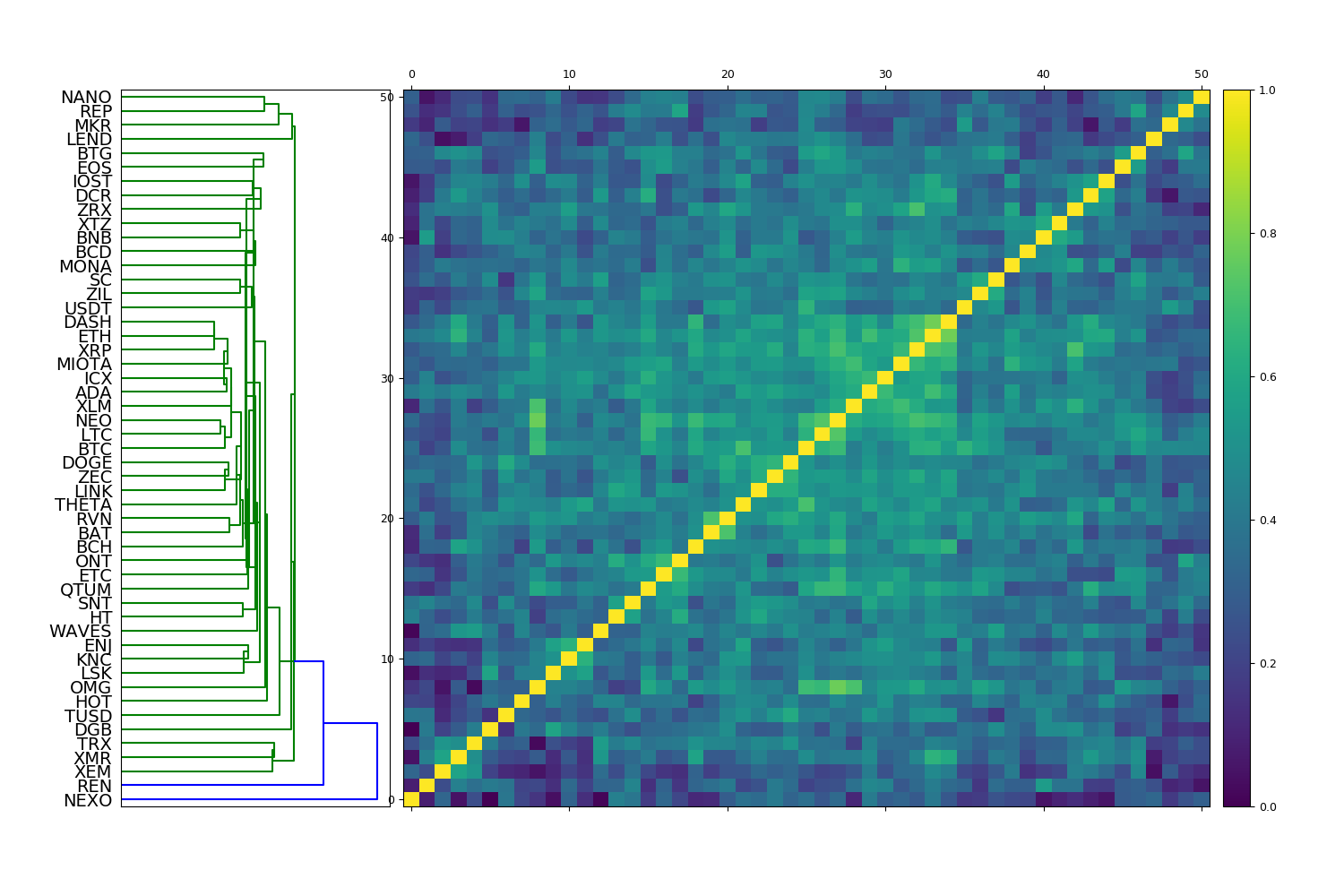}
        \caption{$A_{\text{pre}}^{BV}$ dendrogram}
    \label{fig:Volatility_breaks_dendrogram_pre}
    \end{subfigure}
    \begin{subfigure}[b]{0.95\textwidth}
        \includegraphics[width=\textwidth]{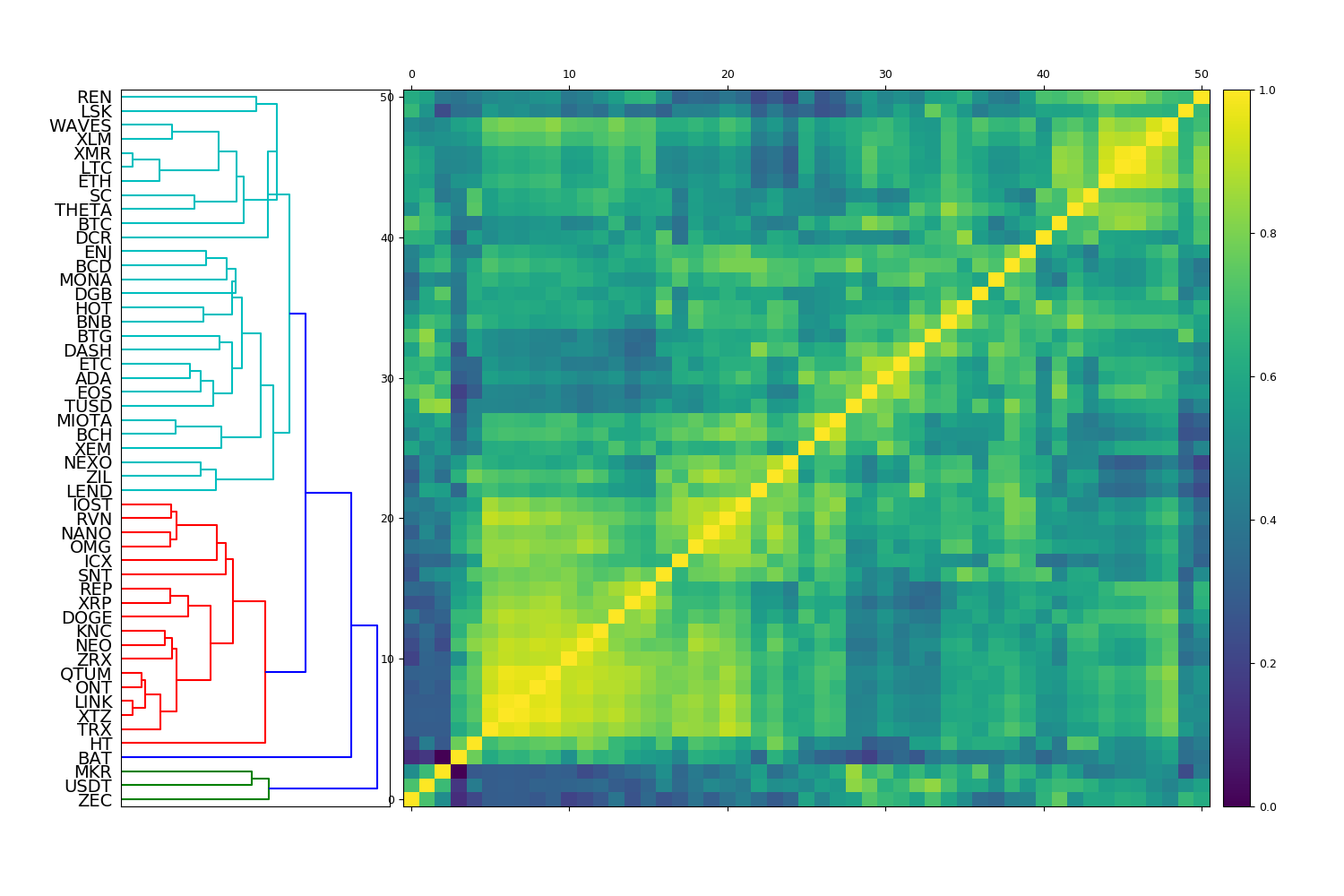}
        \caption{$A_{\text{post}}^{BV}$ dendrogram}
    \label{fig:Volatility_breaks_dendrogram_post}
    \end{subfigure}
    \caption{Hierarchical clustering of affinity matrix between structural breaks with respect to Parkinson variance for (a) pre-COVID and (b) post-COVID.}
    \label{fig:VolatilitybreaksDendrogram}
\end{figure*}

\subsection{Distribution extremities with respect to log returns}
\begin{figure}
    \centering
    \begin{subfigure}[b]{0.95\textwidth}
        \includegraphics[width=\textwidth]{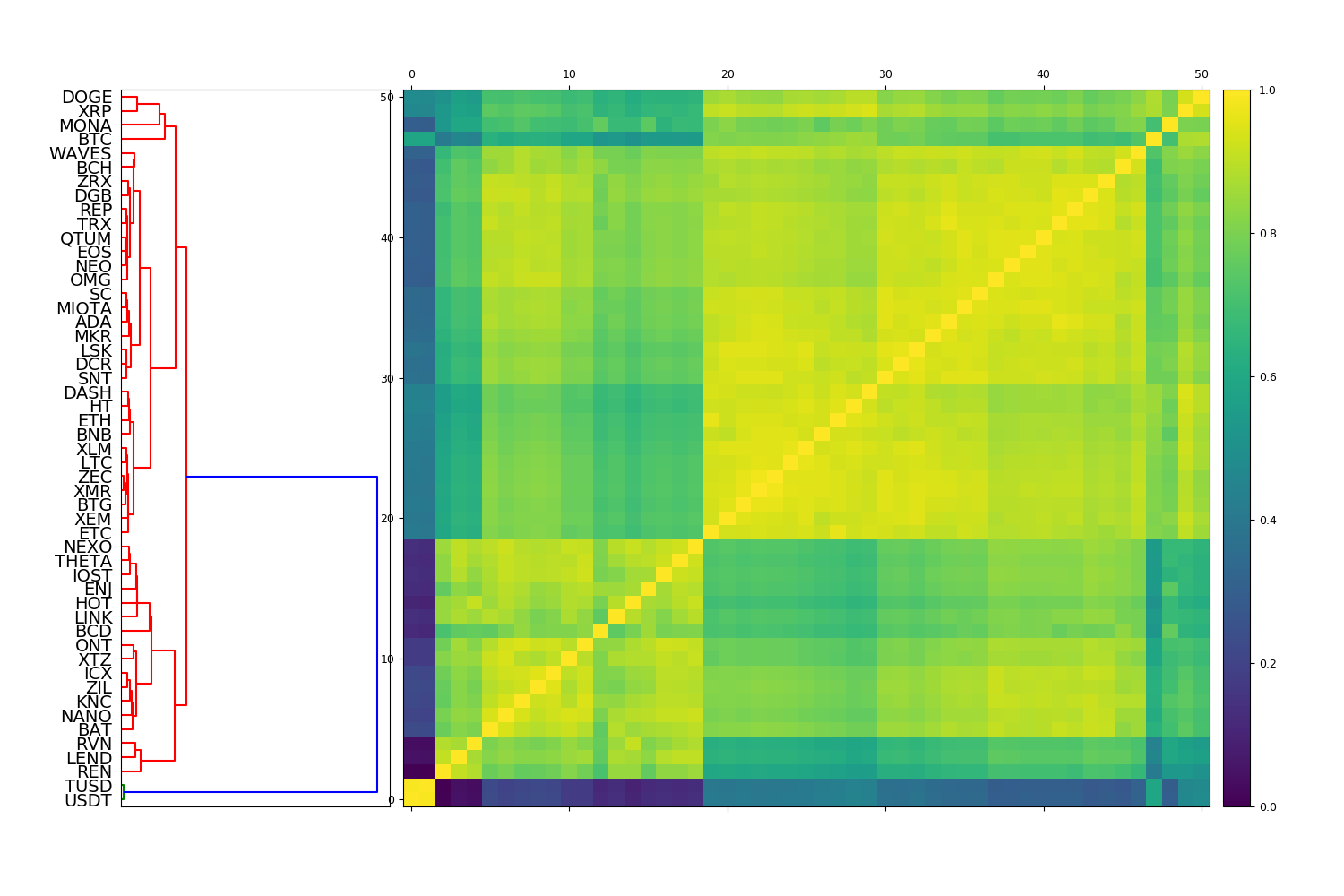}
        \caption{$A_{\text{pre}}^{ER}$ dendrogram outliers}
    \label{fig:Return_extreme_dendrogram_pre}
    \end{subfigure}
    \begin{subfigure}[b]{0.95\textwidth}
        \includegraphics[width=\textwidth]{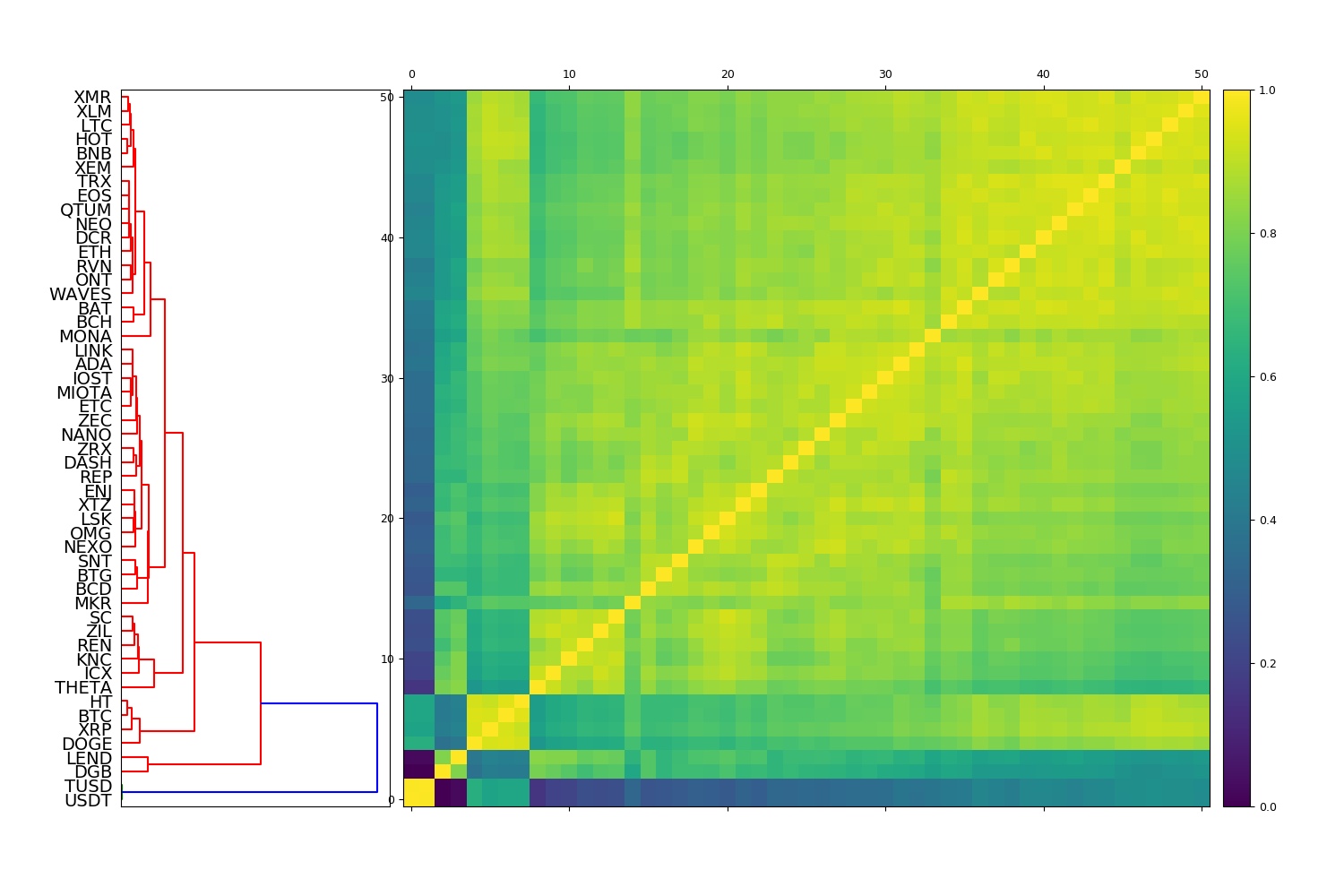}
        \caption{$A_{\text{post}}^{ER}$ dendrogram outliers}
    \label{fig:Return_extreme_dendrogram_post}
    \end{subfigure}
    \caption{Hierarchical clustering of affinity matrix between distribution extremities with respect to log returns for (a) pre-COVID, (b) post-COVID.}
    \label{fig:ReturnsExtremesDendrogram}
\end{figure}

\begin{figure}
    \centering
    \begin{subfigure}[b]{0.45\textwidth}
        \includegraphics[width=\textwidth]{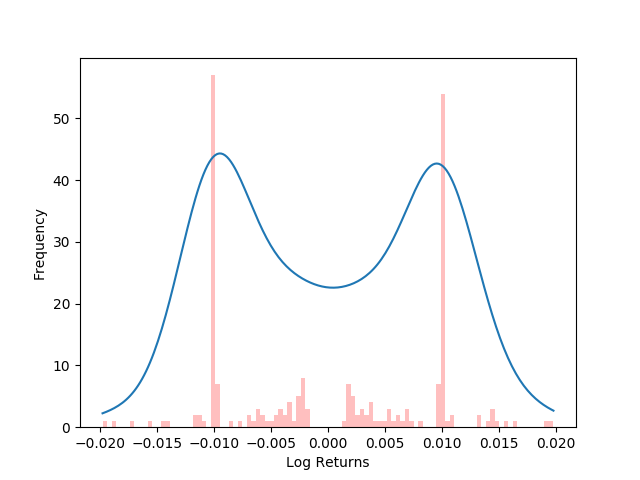}
        \caption{USDT}
    \label{fig:USDT_pre_covid}
    \end{subfigure}
    \begin{subfigure}[b]{0.45\textwidth}
        \includegraphics[width=\textwidth]{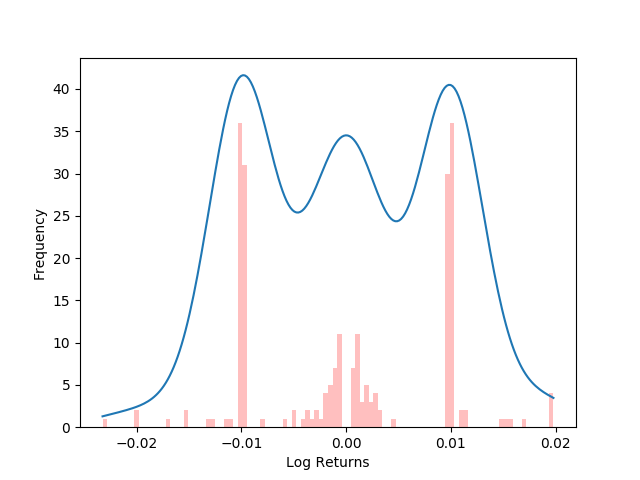}
        \caption{TUSD}
    \label{fig:TUSD_pre_covid}
    \end{subfigure}
        \begin{subfigure}[b]{0.45\textwidth}
        \includegraphics[width=\textwidth]{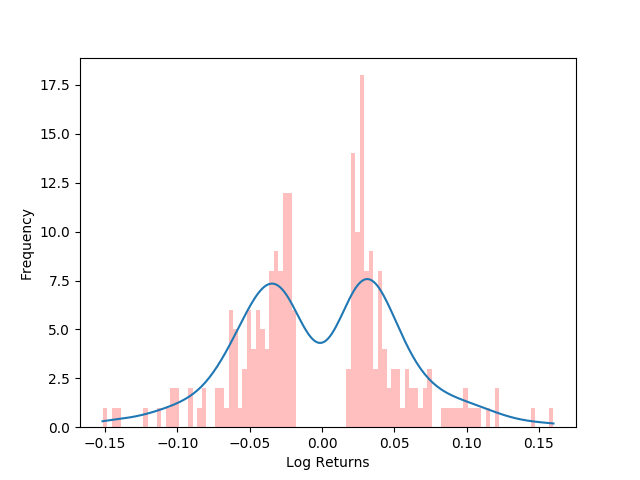}
        \caption{BTC}
    \label{fig:BTC_pre_covid}
    \end{subfigure}
    \begin{subfigure}[b]{0.45\textwidth}
        \includegraphics[width=\textwidth]{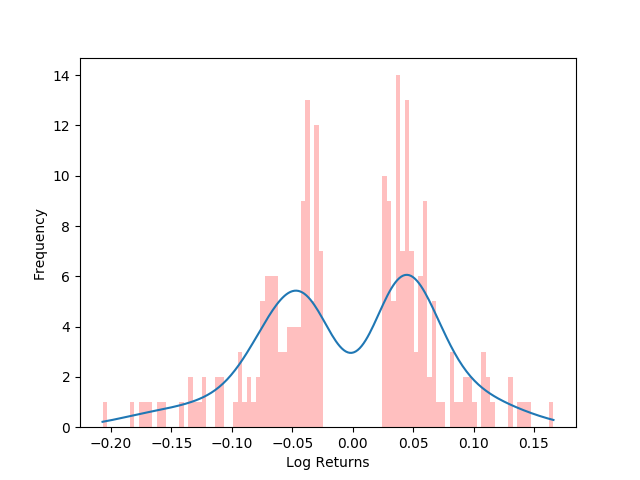}
        \caption{ETH}
    \label{fig:ETH_pre_covid}
    \end{subfigure}
    \begin{subfigure}[b]{0.45\textwidth}
        \includegraphics[width=\textwidth]{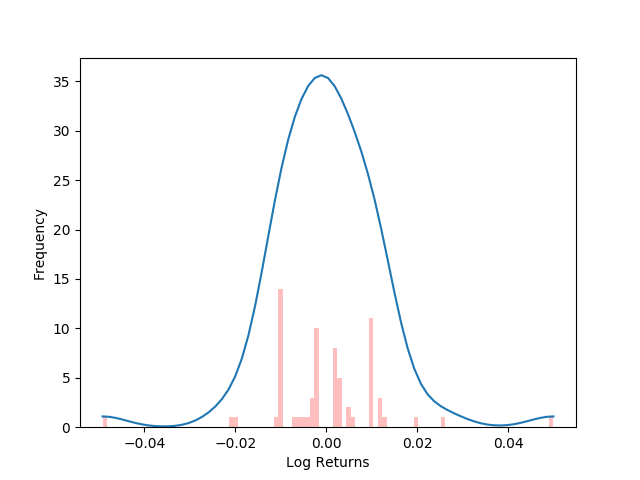}
        \caption{USDT}
    \label{fig:USDT_post_covid}
    \end{subfigure}
    \begin{subfigure}[b]{0.45\textwidth}
        \includegraphics[width=\textwidth]{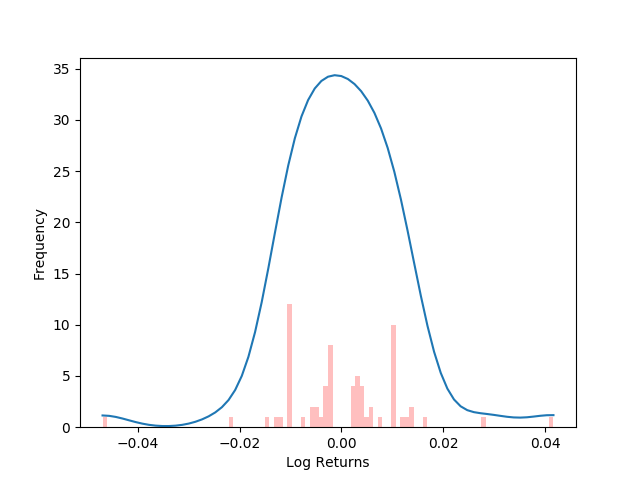}
        \caption{TUSD}
    \label{fig:TUSD_post_covid}
    \end{subfigure}
        \begin{subfigure}[b]{0.45\textwidth}
        \includegraphics[width=\textwidth]{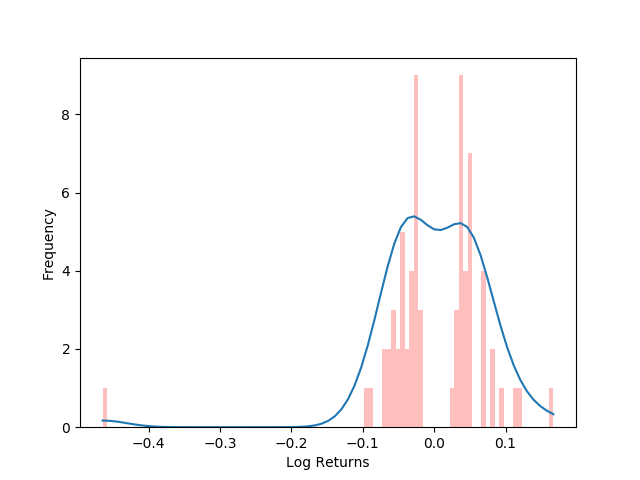}
        \caption{BTC}
    \label{fig:BTC_post_covid}
    \end{subfigure}
    \begin{subfigure}[b]{0.45\textwidth}
        \includegraphics[width=\textwidth]{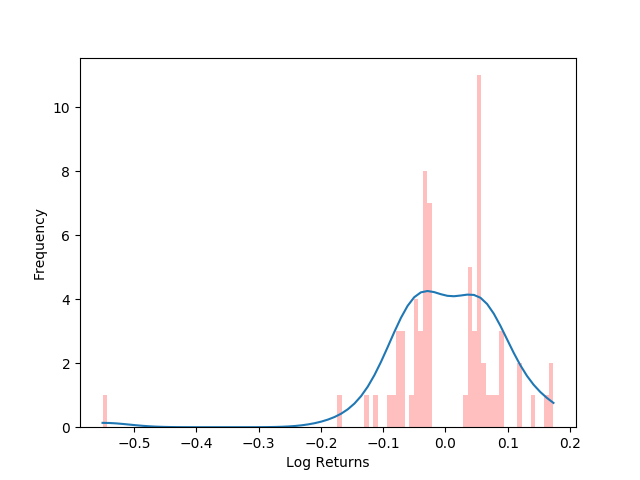}
        \caption{ETH}
    \label{fig:ETH_post_covid}
    \end{subfigure}
    \caption{Extremities of log returns distributions, together with kernel density estimation plots. Figures (a)-(d) represent pre-COVID distributions, (e)-(h) represent post-COVID distributions.}
    \label{fig:Returns_extreme_distributions}
\end{figure}

\begin{figure}
    \centering
    \begin{subfigure}[b]{0.95\textwidth}
        \includegraphics[width=\textwidth]{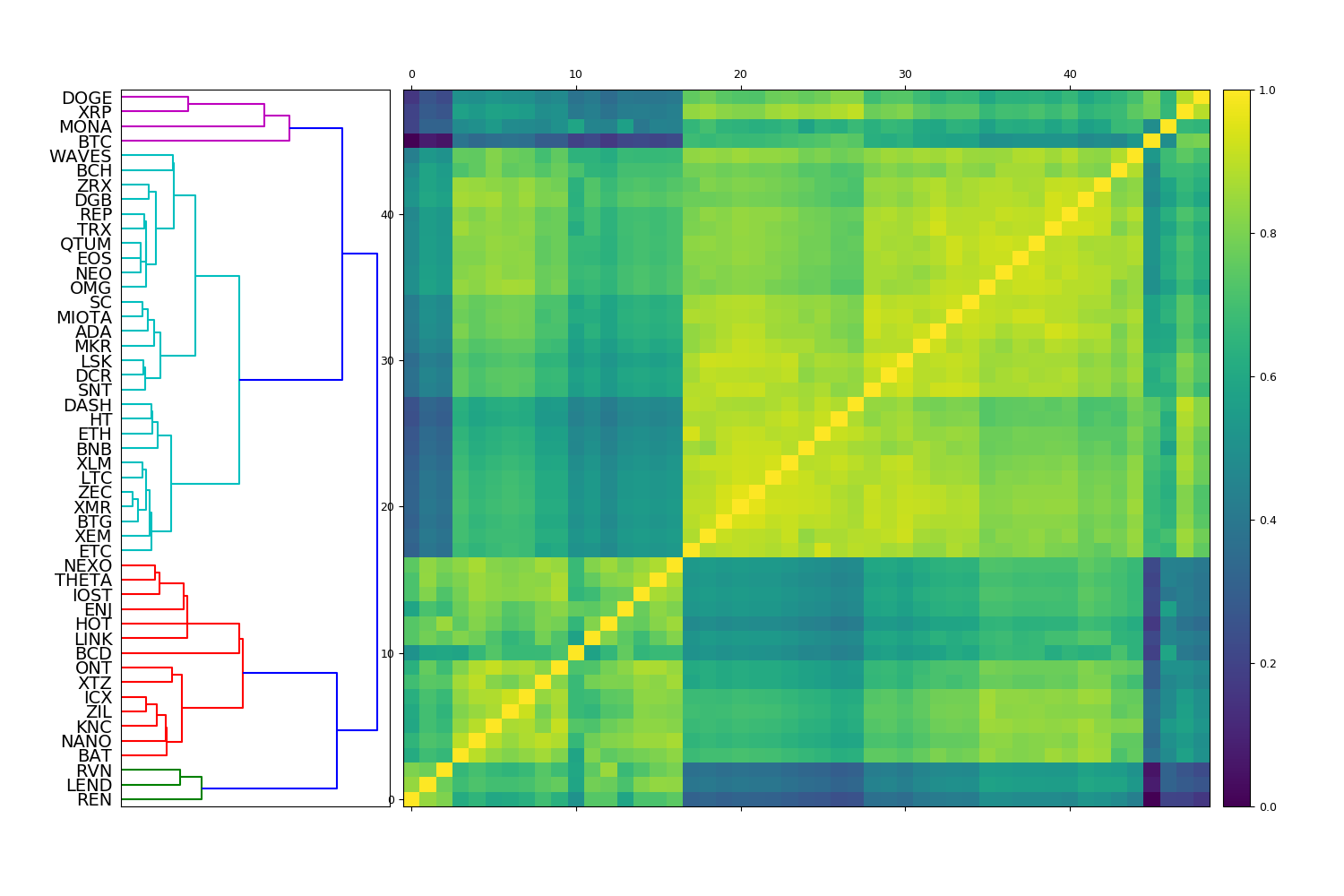}
        \caption{$A_{\text{pre}}^{ER}$ dendrogram no outliers}
    \label{fig:Clean_return_extreme_dendrogram_pre}
    \end{subfigure}
    \begin{subfigure}[b]{0.95\textwidth}
        \includegraphics[width=\textwidth]{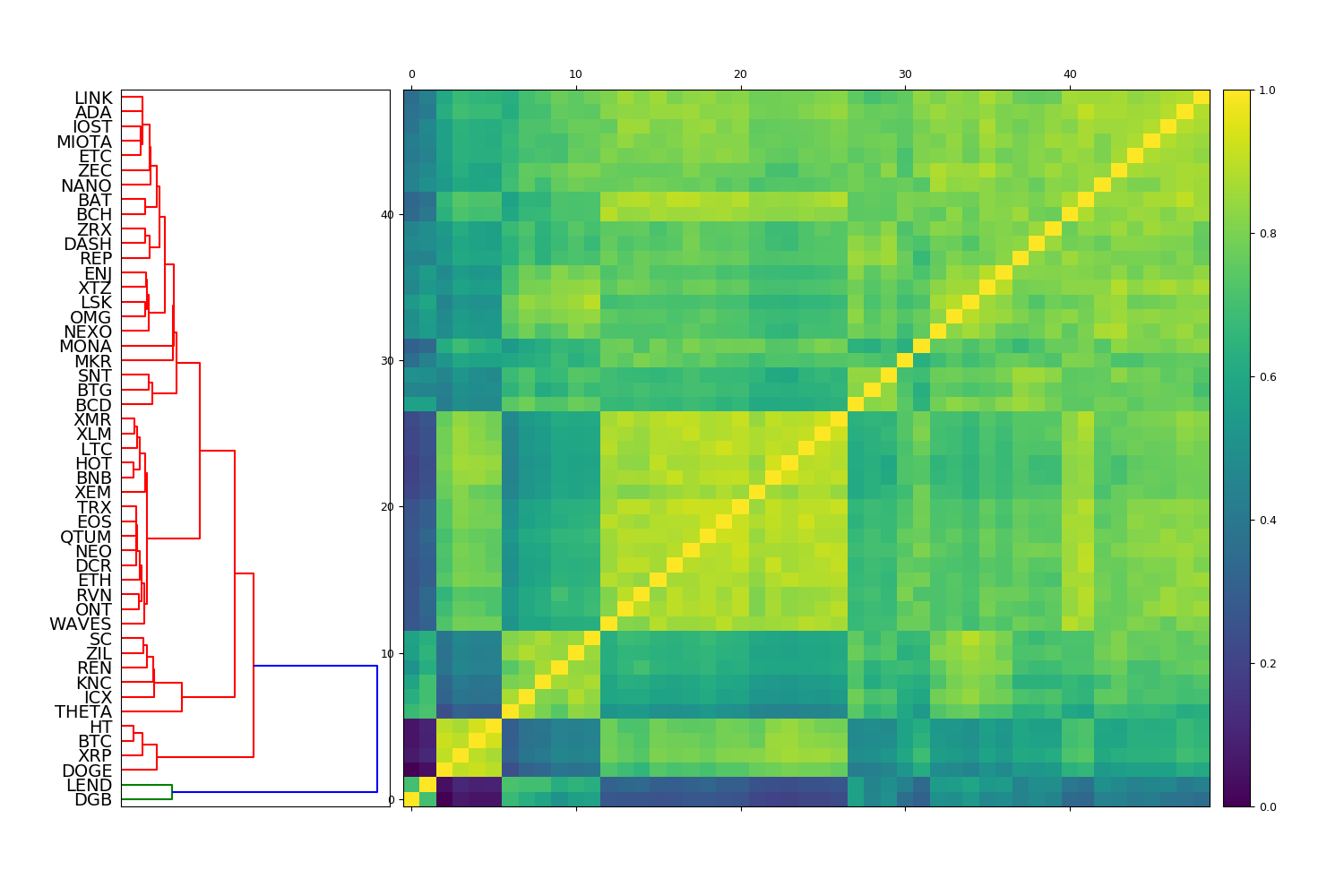}
        \caption{$A_{\text{post}}^{ER}$ dendrogram no outliers}
    \label{fig:Clean_return_extremes_dendrogram_post}
    \end{subfigure}
    \caption{Hierarchical clustering of affinity matrix between distribution extremities with respect to log returns for (a) pre-COVID and (b) post-COVID, excluding outliers USDT and TUSD.}
    \label{fig:CleanReturnsExtremesDendrogram}
\end{figure}

In this section, we analyse the impact of COVID-19 on the distribution extremities of the log returns. Analysis of the corresponding affinity matrices $A_{\text{pre}}^{ER}$ and $A_{\text{post}}^{ER}$ reveals more pronounced structure after the emergence of COVID-19. Hierarchical clustering of these matrices is displayed in Figures \ref{fig:Return_extreme_dendrogram_pre} and \ref{fig:Return_extreme_dendrogram_post}. These dendrograms each highlight two outliers, Tether (USDT) and TrueUSD (TUSD), consistent before and during COVID-19. Both these cryptocurrencies are highly anomalous with respect to their log returns distribution extremities, which is likely due to their low volume of trading and their being pegged to the US dollar in a one-to-one ratio. 


Indeed, in Figure \ref{fig:Returns_extreme_distributions}, we depict the distribution extremities of their log returns alongside two cryptocurrencies in the majority cluster, Bitcoin (BTC) and Ethereum (ETH). Comparing Figures \ref{fig:USDT_pre_covid} and \ref{fig:TUSD_pre_covid} with Figures \ref{fig:BTC_pre_covid} and \ref{fig:ETH_pre_covid}, we see almost a tenfold difference in scale in the log returns. Similar behaviour occurs after the emergence of COVID-19 in Figures \ref{fig:USDT_post_covid}, \ref{fig:TUSD_post_covid}, \ref{fig:BTC_post_covid} and \ref{fig:ETH_post_covid}. Moreover, USDT and TUSD's distributions in \ref{fig:USDT_post_covid} and \ref{fig:TUSD_post_covid}, respectively, lack the asymmetry and negative skew observed for BTC and ETH in \ref{fig:BTC_post_covid} and \ref{fig:ETH_post_covid}. The distributions of BTC and ETH, which are representative of the extreme return behaviours of most cryptocurrencies, highlight the negative skew securities usually experience during times of market crisis.

We note that USDT and TUSD did not stand out as clear outliers, nor did they cluster together, in Figure \ref{fig:VolatilitybreaksDendrogram}, the focus of Section \ref{sec: breaks variance}. Although USDT and TUSD's price movements are highly anomalous and similar to each other in scale, which distinguishes them as outliers in their distribution extremities, the same does not hold for their structural breaks. Indeed, most change point algorithms, including the one that we implement, are independent of scale. They seek to identify points at which the properties of a given time series change. We apply the same change point algorithm independently to each cryptocurrency, so there is no reason that the two cryptocurrencies with the narrowest scale should have similar structural breaks. This shows that TUSD and USDT, while moving much more narrowly than other currencies and hence exhibiting much less extreme behaviour, do not necessarily exhibit similar erratic behaviour.

Next, we treat USDT and TUSD as outliers and analyse the collection after removing them, displaying this subset in Figures \ref{fig:Clean_return_extreme_dendrogram_pre} and \ref{fig:Clean_return_extremes_dendrogram_post}. Figure \ref{fig:Clean_return_extreme_dendrogram_pre} determines four distinct clusters, two of which contain the majority of cryptocurrencies. In Figure  \ref{fig:Clean_return_extremes_dendrogram_post}, which depicts the post-COVID period, all but two cryptocurrencies, Aave (LEND) and DigiByte (DGB), belong to one dominant cluster, which contains several subclusters of high similarity. The difference in the number of clusters and structure of these dendrograms highlights the significant impact COVID-19 has had on extreme return behaviours.

Next, we take a closer look at the cluster structure of the return extremes. First we note that USDT and TUSD were removed from the collection due to their considerably lesser scale in distribution extremities. Statistically, this reflects their much smaller variance and kurtosis than all other cryptocurrencies. To study the remainder of the collection, which have similar scale in their distribution extremities, we use the mean of these restricted distributions instead, as defined in Section \ref{methodology:extreme}. We see that the means have a significant relationship with the cluster structure of the sub-collection in both periods:
\begin{enumerate}
    \item Before COVID-19, Figure \ref{fig:Clean_return_extreme_dendrogram_pre} identifies four distinct clusters. After inspection of the dendrogram and Figure \ref{fig:Return_extreme_dendrogram_pre}, we see that these four naturally fit into two larger clusters: one containing BTC and ETH among others, and one containing HOT and LEND, among others. Computing the means of these two clusters reveals a relationship: the four largest values of $\E(\nu_i)$ all lie in the second cluster, and the average of the means $\E_i \E(\nu_i)$ is greater than the mean of the first cluster under a simple statistical test, with $p<0.05.$ Such measures of statistical significance are not necessarily dispositive  of a meaningful relationship \cite{Amrhein2019}, but ordering the means of the restricted distributions shows that the latter cluster contains more restricted distributions with positive mean, including the greatest four.
     
    \item During COVID-19, excluding outliers USDT and TUSD, the two anomalous cryptocurrencies LEND and DGB have the two greatest means $\E(\nu_i)$ of the entire post-COVID collection. That is, the mean $\E(\nu_i)$ is once again significantly related to the cluster structure.
    \item Before COVID-19, 88\% of cryptocurrencies had an negative value of $\E(\nu_i)$. After the emergence of COVID-19, just 37\% of cryptocurrencies had a negative mean.  That is, a majority of post-COVID return extremes had a positive mean. This is an unexpected result and demonstrates that although there was a sharp initial drop in returns due to COVID-19, this behaviour was anomalous relative to the rest of the post-COVID period. 
\end{enumerate}

\subsection{Identification of cryptocurrency anomalies}
In this section, we focus on identifying two types of anomalies within the cryptocurrency market - cryptocurrencies that are inconsistent between extreme and erratic behaviour, or inconsistent with respect to time. As defined in Section \ref{methodology:crosscontext}, we use the shorthand \emph{inconsistent with respect to behaviour} to mean that a cryptocurrency is irregular with respect to the rest of the market in the comparison between extreme and erratic behaviour, and detect this through our \emph{behaviour inconsistency matrices}, also defined in Section \ref{methodology:crosscontext}. Similarly, we have the shorthand \emph{inconsistent with respect to time}, which describes cryptocurrencies whose behaviour changed significantly in the period of COVID-19. Each inconsistency matrix assigns anomaly scores, which can rank the most inconsistent cryptocurrencies in either regard. We list the top 3 most inconsistent cryptocurrencies pertaining to a range of inconsistency matrices. We have chosen to list the top 3 for ease of interpreting the results, and to point out that examining the most inconsistent cryptocurrencies reveals a considerable number of repetitions - that is, certain cryptocurrencies that are inconsistent in one regard frequently are inconsistent in other regards.

First, we study the behaviour inconsistency matrices $\text{INC}^{\text{pre},R}_{\text{bhvr}}, \text{INC}^{\text{pre},V}_{\text{bhvr}},$  $\text{INC}^{\text{post},R}_{\text{bhvr}}, \text{INC}^{\text{post},V}_{\text{bhvr}}$. We implement hierarchical clustering in Figures \ref{fig:ReturnsTailConsistencyMatrixDendrograms} and \ref{fig:VarianceTailConsistencyMatrixDendrograms}  to highlight clusters in returns and variance, and compute anomaly scores. The behaviour inconsistency dendrogram for pre-COVID returns is displayed in Figure \ref{fig:return_consistency_dendrogram_pre} and reveals a separate cluster of Holo (HOT), 0x (ZRX) and Aave (LEND). Computing anomaly scores reveals that the top 3 most inconsistent cryptocurrencies with respect to differing behaviour are HOT, ZRX and Maker (MKR) - LEND is fourth. The anomaly scores may give slightly different results than the hierarchical clustering because the anomaly scores feature absolute values to calculate absolute differences with other elements, while the dendrograms cluster based on the positive or negative values of the inconsistency matrix. Rows or columns with negative entries may indicate cryptocurrencies of similar behaviour that should be clustered together, but might not necessarily distinguish the most absolutely inconsistent entries of the collection. The behaviour inconsistency matrix with respect to post-COVID returns is clustered in Figure \ref{fig:return_consistency_dendrogram_post}. NEXO (NEXO), USDT and TUSD form their own cluster; the top 3 anomaly scores are of NEXO, THETA (THETA) and MKR.

When analysing inconsistency matrices between extreme and erratic behaviour of variance, we observe the repetition of inconsistent cryptocurrencies - the top 3 anomaly scores for the pre-COVID variance behaviour inconsistency matrix are of NEM (XEM), NEXO and MKR, while MonaCoin (MONA), Ravencoin (RVN) and Bitcoin Diamond (BCD) form a separate cluster. Observing the dendrograms in Figures \ref{fig:variance_consistency_dendrogram_pre} and \ref{fig:variance_consistency_dendrogram_post} side by side, slightly more structure is observed in the post-COVID period, with a growth in the total number of clusters. USDT and TUSD emerge as inconsistent in their behaviour after the emergence of COVID-19. Indeed, before COVID-19, both their distribution extremities and structural breaks are outliers, leading to relative consistency between the two behaviours, while during COVID-19, their structural breaks in variance become similar to the rest of the market.

Time inconsistency matrices are clustered in Figures \ref{fig:CrosstimeDendrogramExtremes} and \ref{fig:CrosstimeDendrogramBreaks} and anomaly scores computed again. Dendrograms for returns and variance extremes, displayed in Figures \ref{fig:crosstime_dendrogram_return_extremes} and \ref{fig:crosstime_dendrogram_variance_extremes}, are dominated by one cluster - with each inconsistency matrix producing one clearly anomalous cryptocurrency, DigiByte (DGBP and BCD respectively. The top 3 anomaly scorers for these inconsistency matrices are DGB, RVN and HOT, and BCD, MKR and DGB, respectively. We see the repetition of inconsistent cryptocurrencies from prior experiments. This is also the case for structural breaks: the top 3 anomalies for structural breaks in returns are USDT, TUSD and HOT, while NEXO and XEM are the second and third most inconsistent cryptocurrencies across time with respect to structural breaks in variance. That is, we see consistent repetition in the cryptocurrencies that are inconsistent between extreme and erratic behaviours and those which were most affected across time by COVID-19.

\begin{figure*}
    \centering
    \begin{subfigure}[b]{0.95\textwidth}
        \includegraphics[width=\textwidth]{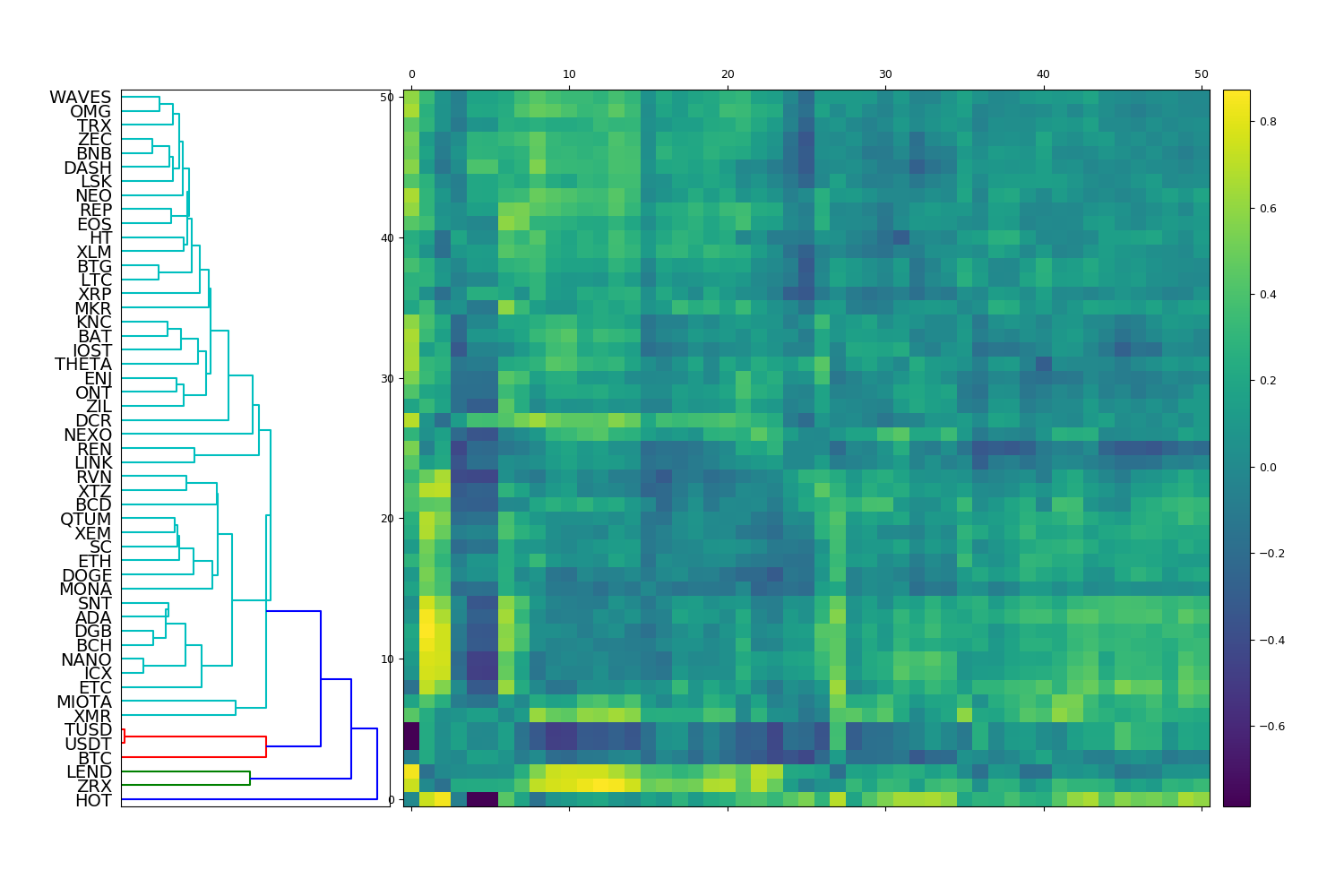}
        \caption{$\text{INC}^{\text{pre},R}_{\text{bhvr}}$}
    \label{fig:return_consistency_dendrogram_pre}
    \end{subfigure}
    \begin{subfigure}[b]{0.95\textwidth}
        \includegraphics[width=\textwidth]{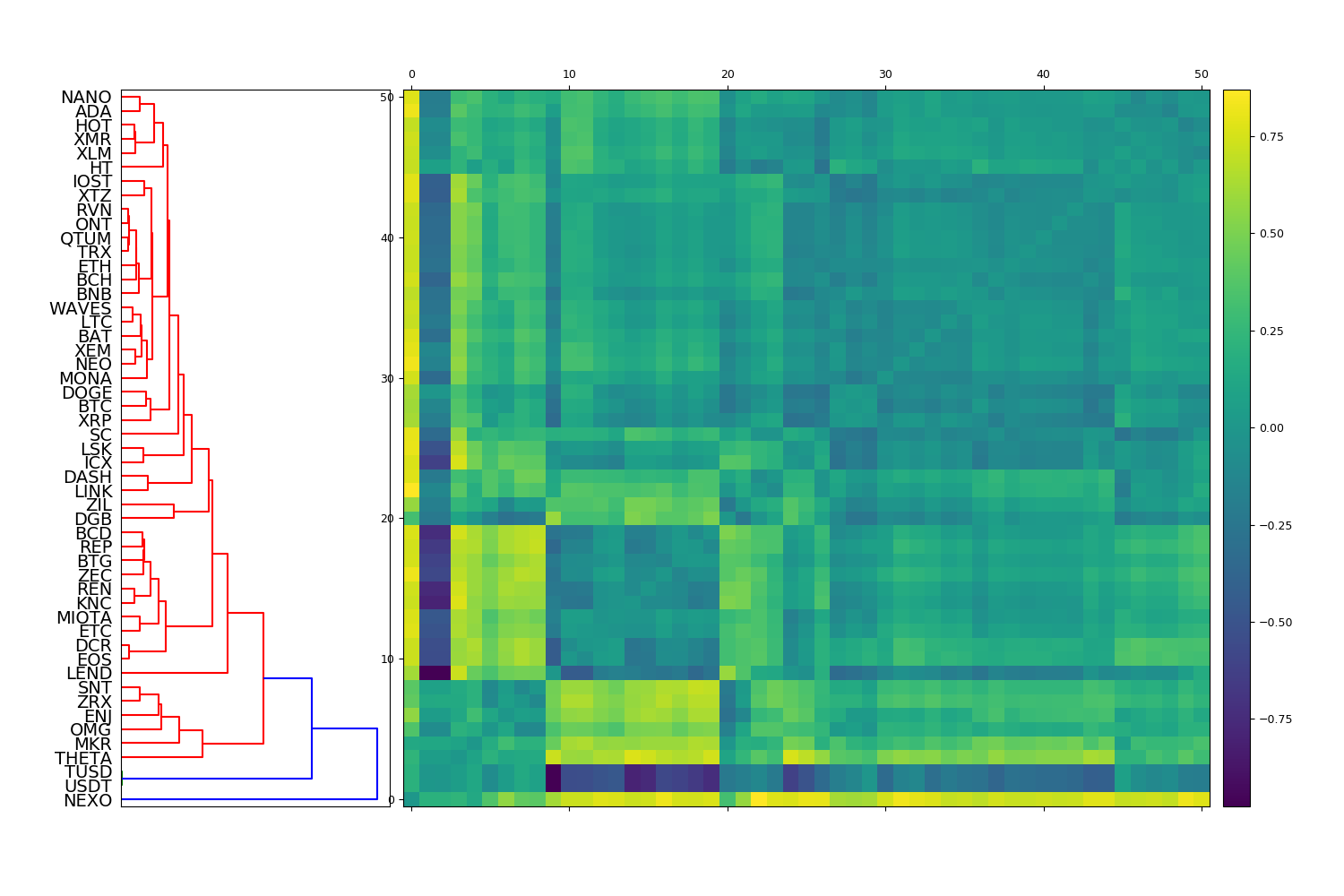}
        \caption{$\text{INC}^{\text{post},R}_{\text{bhvr}}$}
    \label{fig:return_consistency_dendrogram_post}
    \end{subfigure}
        \caption{Hierarchical clustering of inconsistency matrices between extreme and erratic behaviour for (a) log returns pre-COVID (b) log returns post-COVID}
    \label{fig:ReturnsTailConsistencyMatrixDendrograms}
\end{figure*}

\begin{figure*}    
    \begin{subfigure}[b]{0.95\textwidth}
        \includegraphics[width=\textwidth]{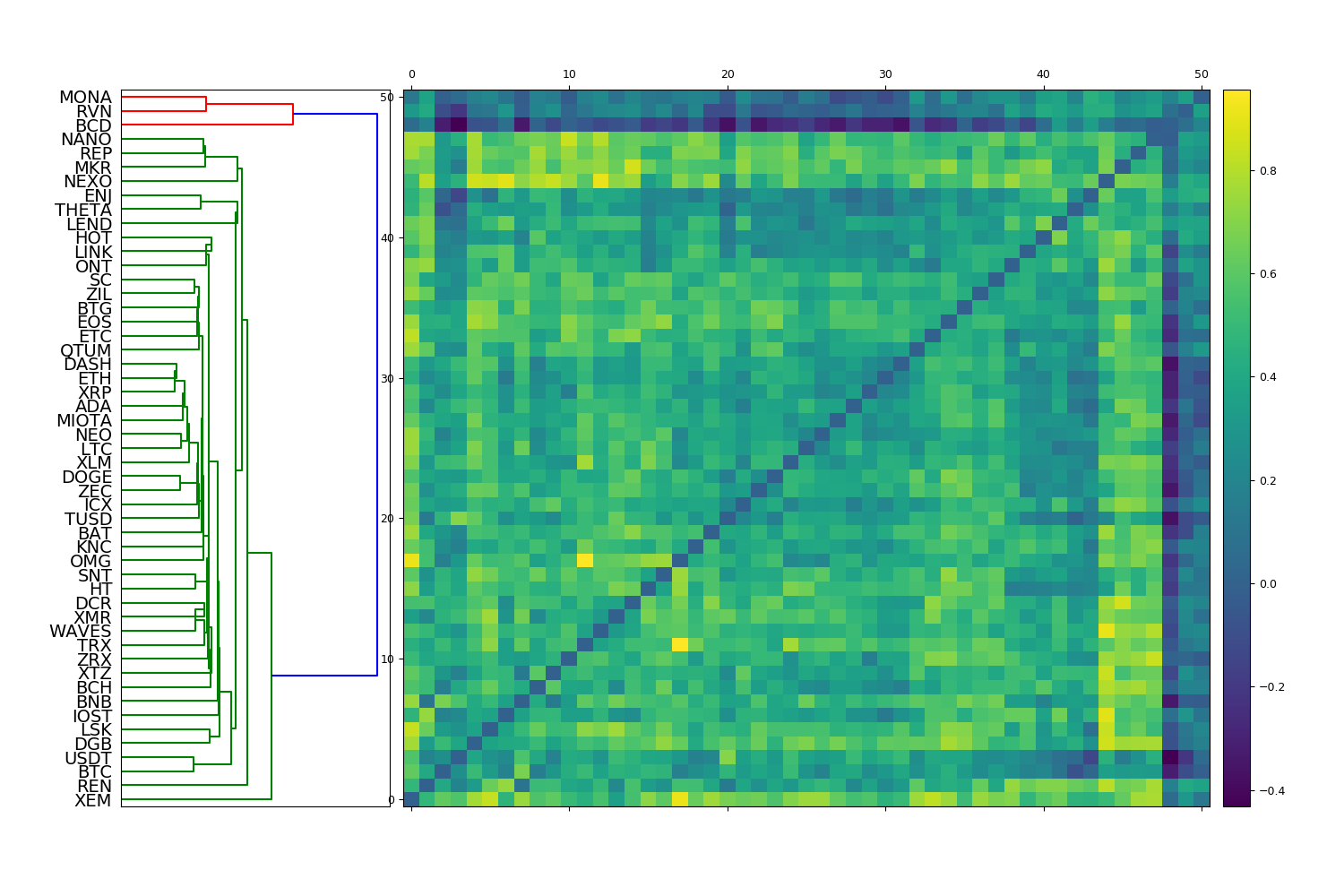}
        \caption{$\text{INC}^{\text{pre},V}_{\text{bhvr}}$}
    \label{fig:variance_consistency_dendrogram_pre}
    \end{subfigure}
    \begin{subfigure}[b]{0.95\textwidth}
        \includegraphics[width=\textwidth]{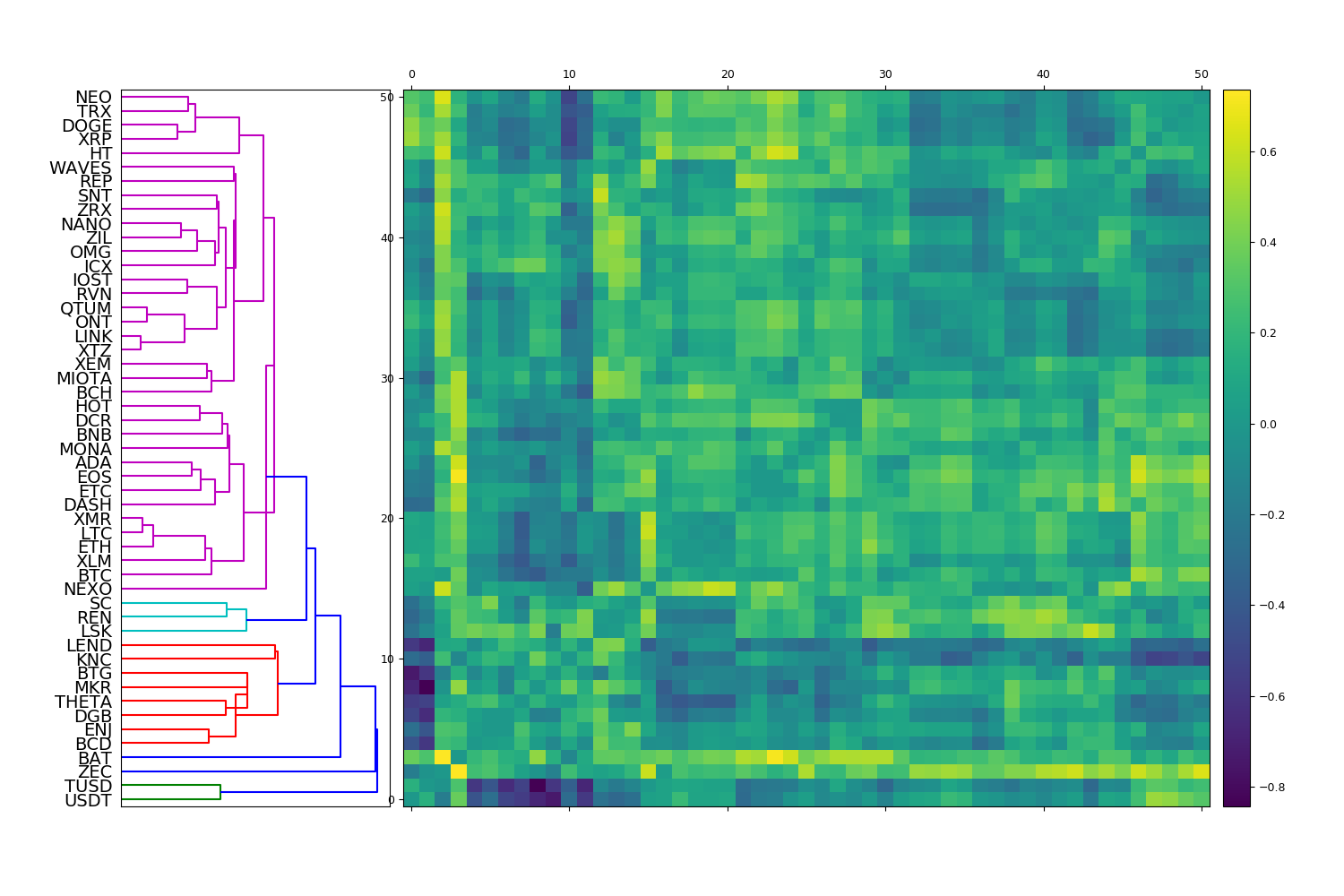}
        \caption{$\text{INC}^{\text{post},V}_{\text{bhvr}}$}
    \label{fig:variance_consistency_dendrogram_post}
    \end{subfigure}
    \caption{Hierarchical clustering of inconsistency matrices between extreme and erratic behaviour for (a) Parkinson variance pre-COVID (b) Parkinson variance post-COVID.}
    \label{fig:VarianceTailConsistencyMatrixDendrograms}
\end{figure*}

\begin{figure*}
    \centering
    \begin{subfigure}[b]{0.95\textwidth}
        \includegraphics[width=\textwidth]{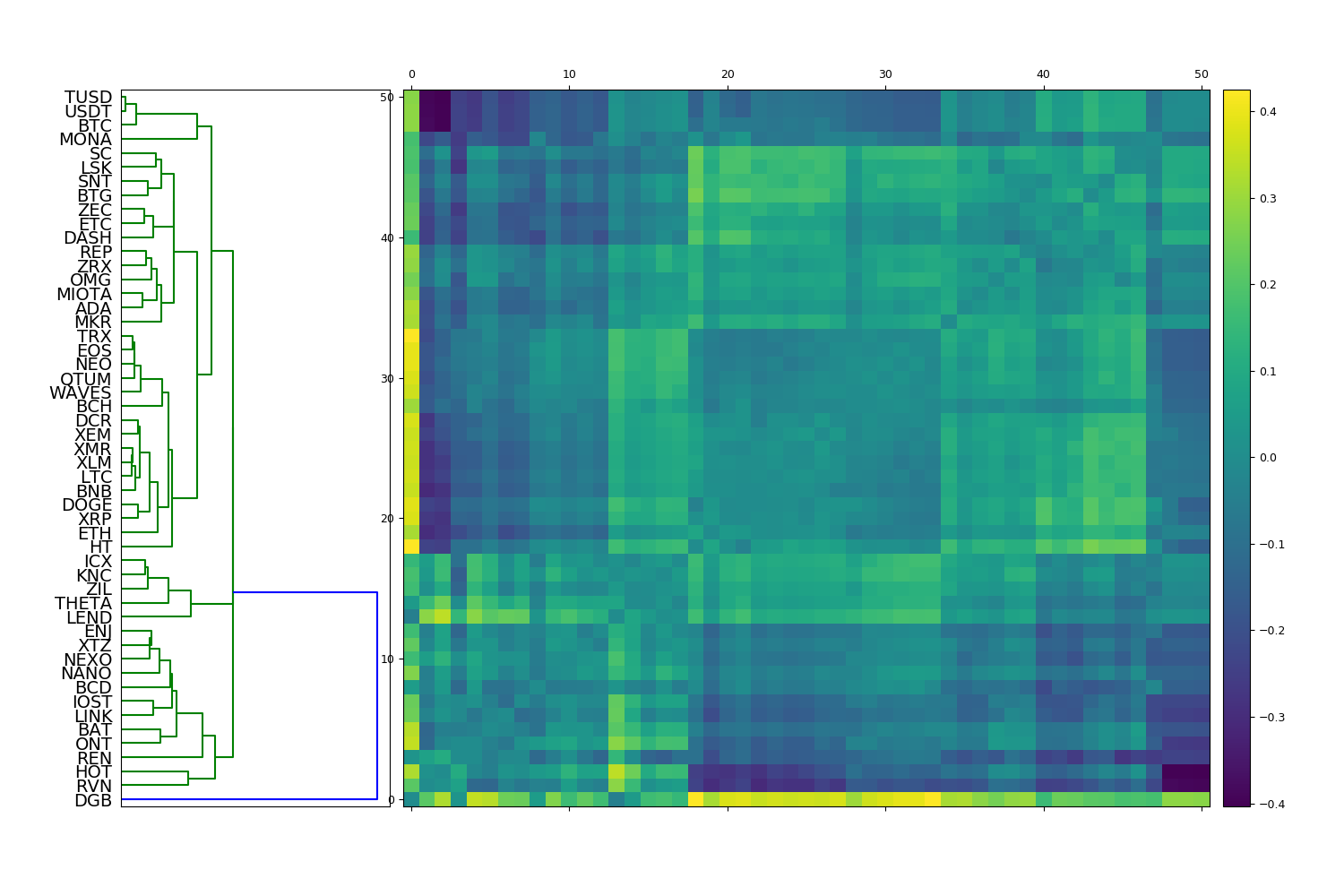}
        \caption{$\text{INC}_{\text{time}}^{ER}$} 
    \label{fig:crosstime_dendrogram_return_extremes}
    \end{subfigure}
    \begin{subfigure}[b]{0.95\textwidth}
        \includegraphics[width=\textwidth]{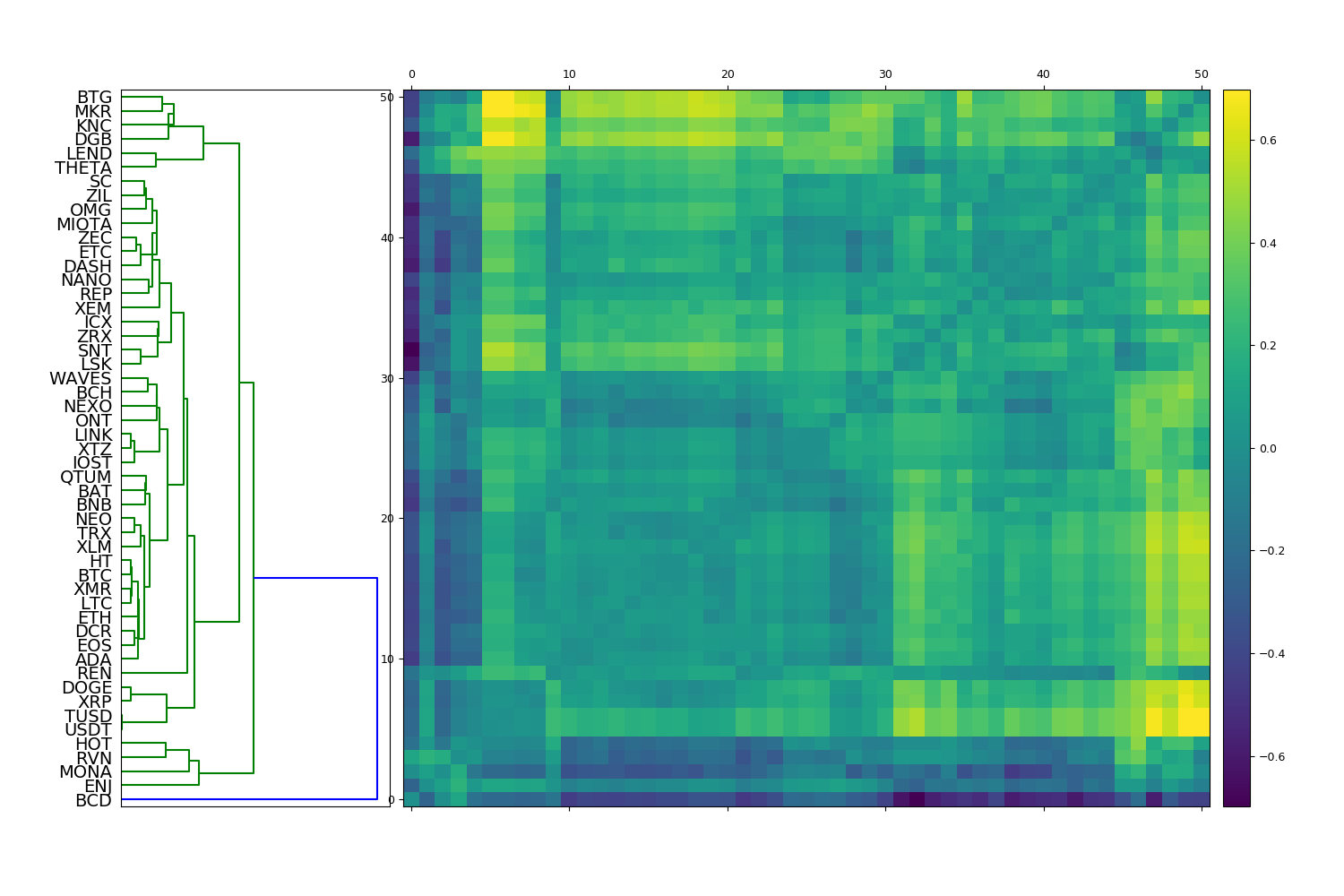}
        \caption{$\text{INC}_{\text{time}}^{EV}$}
    \label{fig:crosstime_dendrogram_variance_extremes}
    \end{subfigure}
    \caption{Hierarchical clustering of time inconsistency matrices for (a) log returns extremes (b) Parkinson variance extremes}
    \label{fig:CrosstimeDendrogramExtremes}
\end{figure*}

\begin{figure*}
    \begin{subfigure}[b]{0.95\textwidth}
        \includegraphics[width=\textwidth]{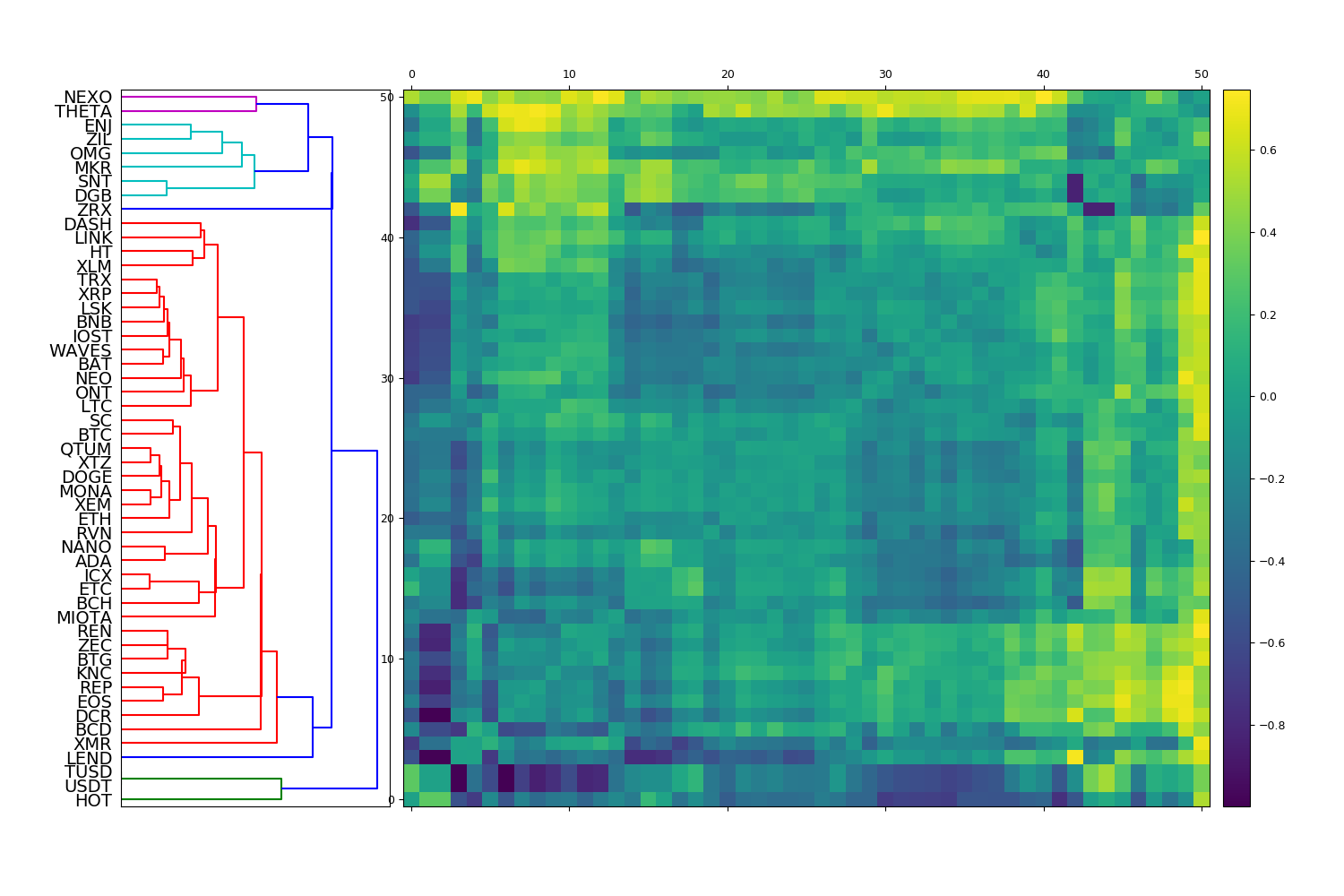}
        \caption{$\text{INC}_{\text{time}}^{BR}$}
    \label{fig:crosstime_dendrogram_return_breaks}
    \end{subfigure}
    \begin{subfigure}[b]{0.95\textwidth}
        \includegraphics[width=\textwidth]{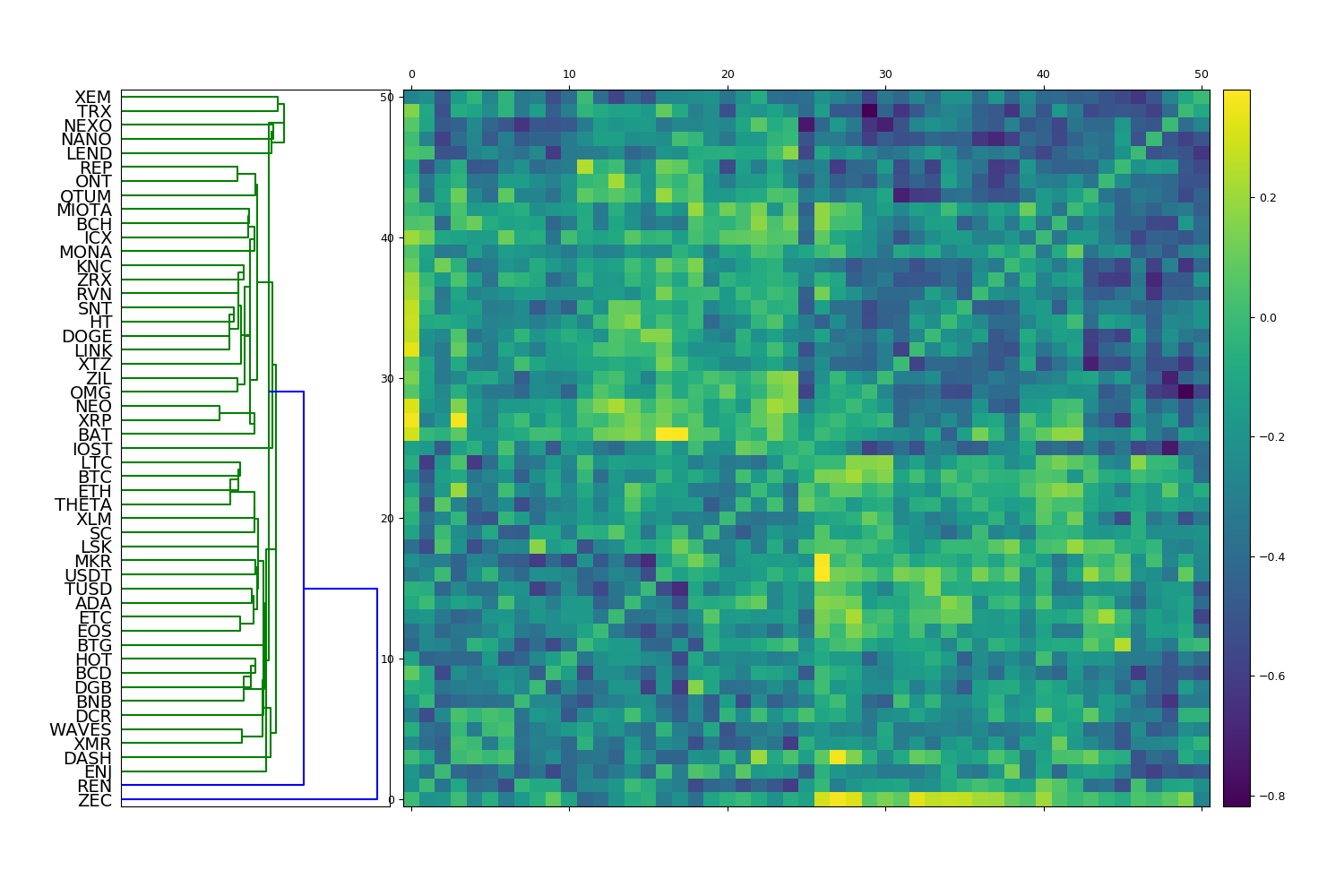}
        \caption{$\text{INC}_{\text{time}}^{BV}$}
    \label{fig:crosstime_dendrogram_variance_breaks}
    \end{subfigure}
    \caption{Hierarchical clustering of time inconsistency matrices for (a) log returns structural breaks (b) Parkinson variance breaks.}
    \label{fig:CrosstimeDendrogramBreaks}
\end{figure*}

\section{Conclusion}
\label{sec:conclusion}
In this paper, we have developed new methods for the study of extreme and erratic behaviours of time series, both individually and in collections. Our work extends the new distance measures between finite sets proposed in \cite{James2020_nsm}. We introduce inconsistency matrices and anomaly scores for identification of elements that are inconsistent with respect to behaviour and time. Such methodology to study extreme behaviour complements the work of \cite{Qi2019,Yang2018,Yang2019}, who have developed machine learning methods in other applications.

Applied to the cryptocurrency market, we have uncovered several insights and dissimilarities when studying the behaviour patterns of returns and variance. In general, cryptocurrency behaviour is more self-similar in variance than returns, both before and during the pandemic. Before COVID-19, the cryptocurrency market exhibited considerable homogeneity with respect to the structural breaks in variance. This was disrupted by the pandemic, with a reduction in self-similarity, reflected in the comparison of Frobenius norms and the emergence of outliers in hierarchical clustering. COVID-19 also had an impact on the return extremes, with an unexpected shift towards positive average means among the distribution extremities. Relatively little impact on the structural breaks in returns was observed. 

Our analysis highlights several anomalies in extreme and erratic behaviour over time, most notably, USDT and TUSD. Both cryptocurrencies exhibit unusually docile profiles for extreme behaviours, likely due to their thin trading structure and lack of market liquidity. Excluding these, several cryptocurrencies repeatedly appeared to be inconsistent between extreme and erratic behaviour and inconsistent over time, namely HOT, NEXO, MKR and XEM, each of which features in the top 3 anomaly scores for at least one inconsistency matrix with respect to time and behaviour.

The precise methodology and applications described in this paper are not an exhaustive representation of the utility of this method. These methods could be used to study structure and identify anomalies among collections of time series in other applications within nonlinear dynamics and with respect to aspects of nonlinearity other than extreme and erratic behaviour.

\section*{Acknowledgements}
Many thanks to Kerry Chen for helpful comments and edits.

\appendix

\section{Glossary}
We include Tables \ref{tab:CryptocurrencyTickers} and \ref{tab:MathematicalObjects}, glossaries of cryptocurrency tickers and mathematical objects introduced in this paper, respectively. Table \ref{tab:CryptocurrencyTickers} orders cryptocurrency by market capitalisation.

\begin{table}[H]
\begin{tabular}{ |p{1.8cm}|p{3.5cm}|p{1.8cm}|p{3.2cm}|}
 \hline
 \multicolumn{4}{|c|}{\textbf{Cryptocurrency tickers and names}} \\
 \hline
 Ticker & Coin Name & Ticker & Coin Name \\
 \hline
 BTC & Bitcoin & DGB & DigiByte \\
 ETH & Ethereum & ZRX & 0x \\
 USDT & Tether & KNC & Kyber Network \\
 XRP & XRP (Ripple) & OMG & OMG Network \\
 BCH & Bitcoin Cash & THETA & THETA \\
 LTC & Litecoin & REP & Augur \\
 BNB & Binance Coin & ZIL & Ziliqa \\
 EOS & EOS & BTG & Bitcoin Gold \\
 ADA & Cardano & DCR & Decred \\
 XTZ & Tezos & ICX & ICON \\
 LINK & Chainlink & QTUM & Qtum \\
 XLM & Stellar & LEND & Aave \\
 XMR & Monero & TUSD & TrueUSD \\
 TRX & Tron & BCD & Bitcoin Diamond \\
 HT & Huobi Token & ENJ & Enjin Coin \\
 NEO & NEO & LSK & Lisk \\
 ETC & Ethereum Classic & REN & Ren \\
 DASH & Dash & NANO & Nano \\
 MIOTA & IOTA & RVN & Ravencoin \\
 ZEC & Zcash & SC & Syscoin \\
 MKR & Maker & WAVES & Waves \\
 ONT & Ontology & MONA & MonaCoin \\
 BAT & Basic Attention Token & NEXO & Nexo \\
 XEM & NEM & HOT & Holo \\
 DOGE & Dogecoin & IOST & IOST \\
 SNT & Status & & \\
\hline
\end{tabular}
\caption{Cryptocurrency tickers and names}
\label{tab:CryptocurrencyTickers}
\end{table}

\begin{table}[H]
\begin{tabular}{ |p{2.3cm}||p{8.9cm}|}
 \hline
 \multicolumn{2}{|c|}{\textbf{Mathematical objects glossary}} \\
 \hline
 Object & Description \\
 \hline
 $D^{ER}$ & Distance matrix between return extremes \\
    $D^{EV}$ & Distance matrix between variance extremes \\
  $D^{BR}$ & Distance matrix between return structural breaks \\
 $D^{BV}$ & Distance matrix between variance structural breaks \\
 $A^{ER},A^{EV}$, etc & Corresponding affinity matrices \\
$||\mathbf{R}_{t}||$ & Time varying log return Frobenius  vector norm \\
$||\Sigma_{t}||$ & Time varying variance Frobenius vector norm \\
$||D^{\text{BR}}||$ & Frobenius matrix norm \\
 $\text{INC}^{\text{pre},R}_{\text{bhvr}}$ & Pre-COVID returns behaviour inconsistency matrix \\
$\text{INC}^{\text{post},R}_{\text{bhvr}}$ & Post-COVID returns behaviour inconsistency matrix \\
 $\text{INC}^{\text{pre},V}_{\text{bhvr}}$ & Pre-COVID variance behaviour inconsistency matrix \\
$\text{INC}^{\text{post},V}_{\text{bhvr}}$ & Post-COVID variance behaviour inconsistency matrix \\
$\text{INC}_{\text{time}}^{ER}$ & Return extremes time inconsistency matrix \\
$\text{INC}_{\text{time}}^{EV}$ & Variance extremes time inconsistency matrix \\
 $\text{INC}_{\text{time}}^{BR}$ & Return structural breaks time inconsistency matrix \\
$\text{INC}_{\text{time}}^{BV}$ & Variance structural breaks time inconsistency matrix \\
\hline
\end{tabular}
\caption{Mathematical objects and definitions}
\label{tab:MathematicalObjects}
\end{table}

\section{Change point detection algorithm}
\label{Appendix_CPD}
In this section, we provide an outline of change point detection algorithms, and describe the specific algorithm that we implement. Many statistical modelling problems require the identification of \emph{change points} in sequential data. By definition, these are points in time at which the statistical properties of a time series change. The general setup for this problem is the following: a sequence of observations $x_1,x_2,...,x_n$ are drawn from random variables $X_1, X_2,...,X_n$ and undergo an unknown number of changes in distribution at points $\tau_1,...,\tau_m$. One assumes observations are independent and identically distributed between change points, that is, between each change points a random sampling of the distribution is occurring. Following Ross \cite{Ross2015}, we notate this as follows:
\begin{equation*}
    X_{i} \sim 
    \begin{cases}
      F_{0} \text{ if } i \leq \tau_1 \\
      F_{1} \text{ if } \tau_1 < i \leq  \tau_2  \\
      F_{2} \text{ if } \tau_2 < i  \leq \tau_3,  \\
      \hdots
    \end{cases}
\end{equation*}
While this requirement of independence may appear restrictive, dependence can generally be accounted for by modelling the underlying dynamics or drift process, then applying a change point algorithm to the model residuals or one-step-ahead prediction errors, as described by Gustafsson \cite{gustafsson2001}. The change point models applied in this paper follow Ross \cite{Ross2015}.

\subsection{Batch change detection (Phase I)}
This phase of change point detection is retrospective. We are given a fixed length sequence of observations $x_1,\ldots,x_n$ from random variables $X_1,\ldots,X_n$. For simplicity, assume at most one change point exists. If a change point exists at time $k$, observations have a distribution of $F_0$ prior to the change point, and a distribution of $F_1$ proceeding the change point, where $F_0 \neq F_1$. That is, one must test between the following two hypotheses for each $k$: 

\begin{equation*}
    H_0: X_{i} \sim F_0, i = 1,...,n
\end{equation*}

\begin{equation*}
    H_1: X_{i} \sim 
    \begin{cases}
      F_{0} & i = 1,2,...,k \\
      F_{1}, & i = k + 1, k+2, ..., n  \\
    \end{cases}
\end{equation*}
and end with the choice of the most suitable $k$.

One proceeds with a two-sample hypothesis test, where the choice of test is dependent on the assumptions about the underlying distributions. To avoid distributional assumptions, non-parametric tests can be used. Then one appropriately chooses a two-sample test statistic $D_{k,n}$ and a threshold $h_{k,n}$. If $D_{k,n}>h_{k,n}$ then the null hypothesis is rejected and we provisionally assume that a change point has occurred after $x_k$. These test statistics $D_{k,n}$ are normalised to have mean $0$ and variance $1$ and evaluated at all values $1 < k < n$, and the largest value is assumed to be coincident with the existence of our sole change point. That is, the test statistic is then

\begin{equation*}
    D_{n} = \max_{k=2,...,n-1} D_{k,n} = \max_{k=2,...,n-1} \Bigg| \frac{\Tilde{D}_{k,n} - \mu_{\Tilde{D}_{k,n}}}{\sigma_{\Tilde{D}_{k,n}}}  \Bigg|
\end{equation*}
where $\Tilde{D}_{k,n}$ were our unnormalised statistics. This test statistic is known as the \emph{Mann-Whitney test} \cite{Ross2015}.
\\

The null hypothesis of no change is then rejected if $D_{n} > h_n$ for some appropriately chosen
threshold $h_n$. In this circumstance, we conclude that a (unique) change point has occurred and its location is the value of $k$ which maximises $D_{k,n}$. That is,

\begin{equation*}
    \hat{\tau} = \argmax_k D_{k,n}.
\end{equation*}

This threshold $h_n$ is chosen to bound the Type 1 error rate as is standard in statistical hypothesis testing. First, one specifies an acceptable level $\alpha$ for the proportion of false positives, that is, the probability of falsely declaring that a change has occurred if in fact no
change has occurred. Then, $h_n$ should be chosen as the upper $\alpha$ quantile of the distribution
of $D_n$ under the null hypothesis. For the details of computation of this distribution, see \cite{Ross2015}. Computation can often be made easier by taking appropriate choice and storage of the $D_{k,n}$.

\subsection{Sequential change detection (Phase II)}
In this case, the sequence $(x_t)_{t \geq 1}$ does not have a fixed length. New observations are received over time, and multiple change points may be present. Assuming no change point exists so far, this approach treats $x_1, . . . , x_t$ as a fixed length sequence and computes $D_t$ as in phase I. A change is then flagged if $D_t > h_t$ for some appropriately chosen threshold. If no change is detected, the next observation $x_{t+1}$ is brought into the sequence. If a change is detected, the process restarts from the following observation in the sequence. The procedure therefore consists of a repeated sequence of hypothesis tests.

In this sequential setting, $h_t$ is chosen so that the probability of incurring a Type 1 error is constant over time, so that under the null hypothesis of no change, the following holds:

\begin{equation*}
    P(D_1 > h_1) = \alpha,
\end{equation*}

\begin{equation*}
    P(D_t > h_t | D_{t-1} \leq h_{t-1}, ... , D_{1} \leq h_{1}) = \alpha, \ t > 1.
\end{equation*}
In this case, assuming that no change occurs, the average number of observations received before a false positive detection occurs is equal to $\frac{1}{\alpha}$. This quantity is referred to as the average run length, or ARL0. Once again, there are computational difficulties with this conditional distribution and the appropriate values of $h_t$, as detailed in Ross \cite{Ross2015}.

\bibliographystyle{elsarticle-num-names}
\bibliography{References.bib}
\end{nolinenumbers}
\end{document}